# Effect of cationic chemical disorder on defect formation energies in uranium-plutonium mixed oxides

Didier Bathellier[1], Luca Messina[1], Michel Freyss[1], Marjorie Bertolus[1], Thomas Schuler[2], Maylise Nastar[2], Pär Olsson[3], Emeric Bourasseau[1]

[1] CEA, DES, IRESNE, DEC, SESC, LM2C, F-13108 Saint-Paul-Lez-Durance, France.
[2] Université Paris-Saclay, CEA, Service de Recherche de Métallurgie Physique, 91191, Gif-sur-Yvette, France.
[3] KTH Royal Institute of Technology, Nuclear Engineering, Roslagstullsbacken 21, SE-10691, Stockholm, Sweden.

## Abstract

At the atomic scale, uranium-plutonium mixed oxides $(U,Pu)O_2$ are characterized by cationic chemical disorder, which entails that U and Pu cations are randomly distributed on the cation sublattice. In the present work, we study the impact of disorder on point-defect formation energies in $(U,Pu)O_2$ using interatomic-potential and Density Functional Theory (DFT+$U$) calculations. We focus on bound Schottky defects (BSD) that are among the most stable defects in these oxides. As a first step, we estimate the distance $R_D$ around the BSD up to which the local chemical environment significantly affects their formation energy. To this end, we propose an original procedure in which the formation energy is computed for several supercells at varying levels of disorder. We conclude that the first three cation shells around the BSD have a non-negligible influence on their formation energy ($R_D \simeq 7.0$ Å). We apply then a systematic approach to compute the BSD formation energies for all the possible cation configurations on the first and second nearest neighbor shells around the BSD. We show that the formation energy can range in an interval of 0.97 eV, depending on the relative amount of U and Pu neighboring cations. Based on these results, we propose an interaction model that describes the effect of nominal and local composition on the BSD formation energy. Finally, the DFT+$U$ benchmark calculations show a satisfactory agreement for configurations characterized by a U-rich local environment, and a larger mismatch in the case of a Pu-rich one. In summary, this work provides valuable insights on the properties of BSD defects in $(U,Pu)O_2$, and can represent a valid strategy to study point defect properties in disordered compounds.

## I. Introduction

Uranium-plutonium mixed oxides (MOX) $(U,Pu)O_2$ with plutonium content of around 10 % are used as nuclear fuels in nuclear light-water reactors. MOX with high plutonium content (25-30 %) is also the reference fuel for future Generation-IV fast breeder reactors. During operation, nuclear fuels withstand extreme conditions because of fission reactions that can produce point defects in two different ways. Irradiation-induced defects are created by the neutrons emitted upon a fission reaction, and by the recoil energy of the resulting fission products. In addition, fission reactions give off a large amount of heat that increases the fuel temperature. High temperatures produce thermal point defects due to configurational entropy. The role of thermal and irradiation point defects is critical for the transport of chemical species, in particular fission products whose mobility influences significantly the fuel performance during operation in reactor. Therefore, an accurate determination of point defect properties in $(U,Pu)O_2$ is of great concern. Defect formation energies govern the defect concentrations in a nuclear fuel at a given temperature $T$, and their accurate calculation contributes to assess the impact of point defects on microstructure evolution and macroscopic fuel properties. Furthermore, point defects control the diffusion of substitutional chemical species (direct interstitial diffusion mechanism does not require point defects) in a material. At equilibrium, diffusion coefficient calculations require the determination of the activation energy of the diffusion mechanism, which for substitutional species is equal to the sum of the formation of point defects and some effective migration energy related to the considered mechanism. Moreover, the calculation of defect formation energies is a necessary step to get insight on the trapping energy of fission gases in the fuel crystal lattice. Thus, there are many important reasons to accurately determine the formation energies of thermal point defects in $(U,Pu)O_2$.

$(U,Pu)O_2$ is an ionic-covalent solid in which uranium and plutonium atoms carry a positive charge and oxygen atoms a negative one. $(U,Pu)O_2$ crystallizes in a fluorite-like structure: the $O^{2-}$ anions occupy a simple cubic sublattice, and the $U^{4+}$ and $Pu^{4+}$ cations are distributed on a face-centered cubic (*fcc*) sublattice. MOX fuel is a disordered solid solution, which is known to exhibit a solid solubility across the whole compositional range [1]. In addition, its lattice parameter and its melting temperature are found to follow the rule of mixture according to several experimental studies [2]–[5]. These results lead us to consider $(U,Pu)O_2$ as an ideal solid solution. However, more recent studies consider the effect of non-ideal mixing behavior in $(U,Pu)O_2$ [6], [7] or conclude that $(U,Pu)O_2$ cannot be approximated as an ideal mixture for the whole compositional range [8], [9]. In addition, the presence of point defects may affect the ideality of the solution by introducing some local ordering. In an ideal cationic solution $(U,Pu)O_2$, $U^{4+}$ and $Pu^{4+}$ are randomly distributed on the *fcc* sublattice, which is referred to as cationic chemical disorder. This disorder complicates the determination of point defect properties in $(U,Pu)O_2$ that depend on the U/Pu distribution in the local chemical environment around the defect. Namely, in the case of the pure actinide oxides ($UO_2$ and $PuO_2$), a point defect has only one formation energy because there is only one possible chemical environment. In $(U,Pu)O_2$, a point defect can have various formation energies depending on the local distribution of $U^{4+}$ and $Pu^{4+}$ cations around it. The purpose of the present study is to determine whether the cationic chemical disorder around a point defect affects its formation energy, and to assess this influence. Therefore, computing the defect formation energies for all the possible cation configurations around a defect is in principle necessary to characterize defect properties. However, this task is computationally challenging because of the high computational costs required for disordered compounds, for which the configuration space is very large. The goal of the present study is to propose new strategies to address this issue, in order to study defect properties in $(U,Pu)O_2$.

Many different defects can be created in a material; the most elementary ones are interstitials and vacancies. In $UO_2$, positron annihilation spectroscopy measurements coupled with first-principles calculations [10] have shown that the bound Schottky defect (BSD), i.e., a tri-vacancy made of two oxygen vacancies ($V_O$) and one uranium vacancy ($V_U$), is a predominant radiation damage feature [10], [11] and also a favorable site for fission gas trapping [12], [13]. Data on the BSD formation energy in $(U,Pu)O_2$ is very scarce in the literature. Experimental data is available for $UO_2$ [14] but there is no measurement for $(U,Pu)O_2$ to this day to our knowledge. Cheik Njifon [15] computed this property in $(U,Pu)O_2$ using DFT+$U$ [16] (Density Functional Theory + $U$) for various defects including the BSD, but considering only one possible atomic configuration with a fixed Pu content because of the high cost of electronic structure calculations. Cationic chemical disorder was described in this study with 96-atom supercells generated by von Pezold *et al.* [17] using the Special Quasirandom Structure (SQS) method [18]. However, the defect formation energy in one cation environment only is not representative of the effect that cationic chemical disorder can have on this property. Balboa *et al.* [19] determined BSD formation energies by averaging the results obtained in seven different configurations for a given plutonium content using the Cooper-Rushton-Grimes (CRG) empirical interatomic potential (EIP) [20]. Seven supercells containing 2592 atoms were generated by randomly distributing the $U^{4+}$ and $Pu^{4+}$ cations on the *fcc* sublattice. The calculated BSD formation energies were then obtained as the arithmetic energy average over the seven configurations. This study provided a first insight on the BSD formation energies in MOX fuel using interatomic potentials, but the explored portion of configuration space was still very limited.

In the present study, we perform an extensive exploration of the configuration space of local cation environment around the BSD in order to determine the effect of the cationic chemical disorder on its formation energy in $(U,Pu)O_2$. As the influence of the local environment on the defect properties should reasonably decrease with distance, we first determine the cutoff distance $R_D$ beyond which the effect of the cationic chemical disorder on the defect formation energy becomes negligible. This allows us to reduce the amount of configurations to explore. After a short presentation of the method used to compute the BSD formation energy (section II), we explain in section III how we estimate the cutoff distance $R_D$ based on CRG potential calculations [20]. This potential correctly reproduces thermodynamic properties of $(U,Pu)O_2$ [21] and defect properties in $UO_2$ [22]. Then, we apply in section IV a systematic approach to compute the BSD formation energies, with a much broader exploration (compared to previous studies) of the configuration space that allows for an in-depth analysis of the effect of the cationic chemical disorder on the BSD formation energy. We explore all the possible cation configurations in the first and second atomic shells around a BSD and compute the formation energy associated with each configuration with the CRG potentials. This procedure is applied to several MOX compositions ranging from 0% to 100% Pu. Finally, we carry out in section V DFT+$U$ calculations for some selected configurations to benchmark the results obtained with the CRG potential.

## II. Computation method of bound Schottky defect formation energy

Bound Schottky Defects (BSD) in $(U,Pu)O_2$ consist of one cation ($U^{4+}$ or $Pu^{4+}$) and two anion ($O^{2-}$) vacancies that are first-nearest neighbors of the cation vacancy. There are three different BSD configurations depending on the position of the two oxygen vacancies (see Figure 1). BSD1 (resp. BSD2, BSD3) corresponds to the case where the oxygen vacancies are first- (resp. second-, third-) nearest neighbors on the oxygen sublattice. In Figure 2, we show the cation environment around the oxygen vacancies: let us note that in the case of BSD3 and BSD2, six cations are first-nearest neighbors to either one of the two oxygen vacancies, while there are five of them in the case of BSD1.

In the commonly used supercell method, the formation energy of a vacancy-type defect is computed with respect to a reference, given by the energy of the supercell where the vacancy is replaced by a host atom. The usual formula (which is correct for pure materials) reads:

$$E_f^X = E_D - \frac{N-1}{N} E_R^X \quad (2.1)$$

where $E_D$ is the energy of the supercell containing the defect, $E_R^X$ the energy of the reference supercell, $X$ the host species placed in the vacant site, and $N$ the number of atoms in the supercell without the defect (*i.e.*, the number of cations in the case of cation vacancies in oxides). This is the equation used, among others, by Cheik Njifon [3] and Balboa *et al.* [19] to determine the BSD formation energies in (U,Pu)O₂. Cheik Njifon considered the cases of a U or Pu atom replacing the cation vacancy as two different types of defect (called BSD-U and BSD-Pu, respectively), and computed their formation energies in one SQS-generated cation configuration only. On the other hand, Balboa considered seven different configurations, and computed an average $E_f$ regardless of whether the chosen reference supercell contained a U or a Pu atom in place of the vacancy.

We show in appendix A that this traditional approach given by Eq. 2.1 is not well suited (and rigorously incorrect for the case of a vacancy-type defect in a disordered solution, because the chemical potentials of U and Pu are correctly not accounted for. The energy reference in Eq. 2.1 depends indeed on the nominal Pu concentration in the supercell (see Eq. A6). This leads to formation energies that are not physically meaningful, and hinders the comparison of formation energies obtained in supercells with different Pu concentrations. In addition to this limitation, the concept of distinguishing between BSD-U and BSD-Pu defects is physically unjustified, as the latter are in practice identical defects that should have one equilibrium concentration only, regardless of the chosen reference energy. Note that this is specific to compounds with sublattices occupied by multiple elements, where a given lattice site can be occupied by more than one species.

For these reasons, in this work we rely on a slightly different definition of formation energy, as follows:

$$E_f^X = E_D - E_R^X + \frac{E(XO_2)}{N} \quad (2.2)$$

where $E(XO_2)$ is the energy of the pure, non-defective UO₂/PuO₂ supercells. Note that the difference between Eq. 2.1 and Eq. 2.2 lies in the normalization term: $E(XO_2)/N$ in place of $E_R^X/N$. We show in appendix B that Eq. 2.2 accounts for the chemical potential of each species in a more rigorous way than Eq. 2.1. Namely, the latter is taken equal to that in the bulk system, which is valid under the assumption that the effect of defects or short-range order on the chemical potentials is negligible. In other words, the system with and without defect is treated as an ideal solid solution. In the same appendix, we discuss that choosing a U-based or a Pu-based reference cell is equally valid to analyze the impact of chemical disorder on the defect formation energy, since the difference between $E_f^U$ and $E_f^{Pu}$ extracted from the same surrounding atom configuration is a slight shift depending on the nominal composition [23].

Therefore, we choose to focus on the analysis of $E_f^U$. In Appendix D, we show that the analysis of $E_f^{Pu}$ lead to the same conclusions. We make the assumption that any short-range order effect due to a non-ideality character of the solution (or to the presence of defects) can be neglected. This is the first step of a more general approach where the possible effect of local ordering will be taken into account in future work. The more general formula accounting for short-range order is provided in appendix B (Eqs. B1 and B2).

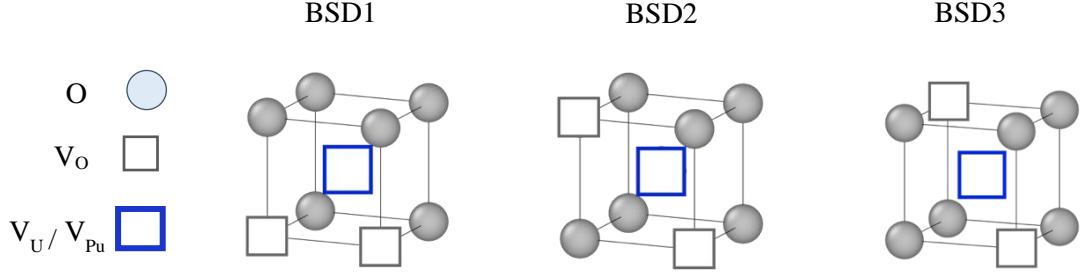

Figure 1: The three different types of Bound Schottky Defects (BSD) in (U,Pu)O$_2$.

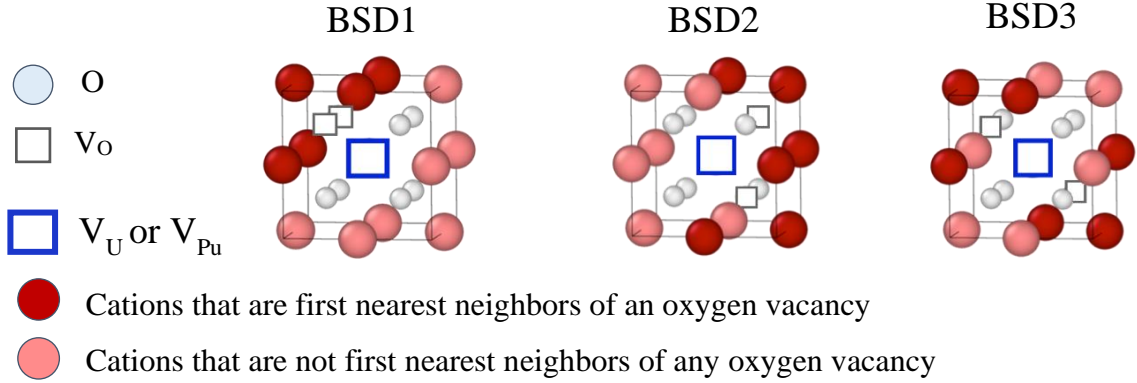

- Cations that are first nearest neighbors of an oxygen vacancy
- Cations that are not first nearest neighbors of any oxygen vacancy

Figure 2: Representation of the neighboring atoms of the oxygen vacancies constituting the BSD1, BSD2 and BSD3 Schottky Defects in (U,Pu)O$_2$.

## III. Approach to estimate the cut-off distance with the CRG potential

As mentioned in the introduction, from a specific cutoff distance $R_D$ around a defect, we can reasonably assume that the cationic chemical disorder does not have a significant effect on the defect formation energy. Therefore, the determination of $R_D$ allows us to reduce the number of configurations to explore to a manageable number. The present section is dedicated to the calculation of $R_D$.

### III.1. Method of generating supercells

To estimate the distance $R_D$ from which the cationic chemical disorder does not significantly modify the BSD formation energy, we compute this property using the Cooper-Rushton-Grimes (CRG) empirical interatomic potential [20], in a large set of 2592-atom supercells (6x6x6). Because of the extremely large size of the configuration space, it is unfeasible to compute the defect formation energy in every possible 2592-atom-supercells of (U,Pu)O$_2$. Thus, we propose to explore the entire range of cationic chemical disorder, i.e., to compute defect formation energies in ordered and disordered supercells, but also in supercells with intermediary cationic disorder. On the one hand, we generate ten disordered supercells containing 50 % Pu using the Special Quasirandom Structure (SQS) method [18]. We denote these ten disordered supercells from SQS 1 to SQS 10. On the other hand, we consider four ordered supercells with a Pu atomic concentration of 50 %. Two of them are organized as an alternation of U and Pu planes that we call "plane supercell 1" and "plane supercell 2" — see Figure 3a and Figure 3b respectively.

The third ordered supercell is an unmixed one where all the uranium atoms are located on one side of the supercell, and all the plutonium atoms on the other side (see Figure 3c). We call this supercell the 1/2-unmixed supercell. The last ordered supercell is separated in eight regions, each of which is composed solely of uranium or solely of plutonium atoms (see Figure 3d). This supercell is denoted as the 1/8-unmixed supercell.

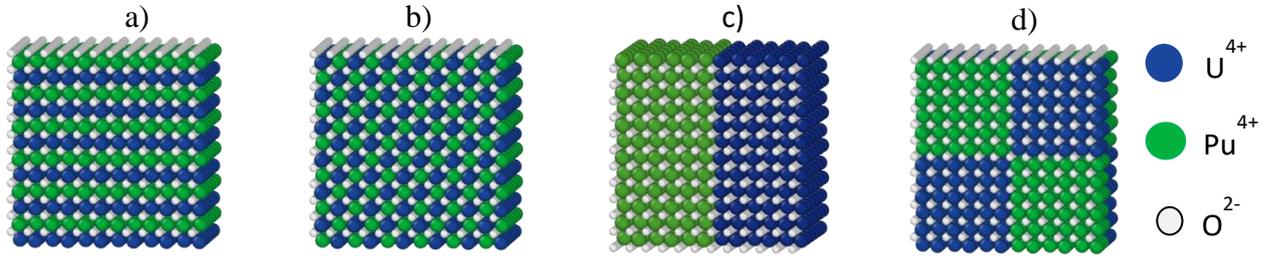

Figure 3: Ordered 2592-atom supercells of (U,Pu)O$_2$ used in the cut-off distance study.

From the generated ordered and SQS supercells, we build additional configurations, which we call composite supercells, using the procedure as follows:

Let us consider a given SQS supercell and a given ordered supercell, for instance SQS 1 and the 1/2-unmixed supercell (Figure 3c).

1. A BSD (BSD1, BSD2 or BSD3) is created at the center of the ordered supercell.
2. The cations first nearest neighbor (1nn) of the cation vacancy of the BSD are removed.
3. The removed cations are substituted by the cations from the SQS supercell that are located at the same lattice position. A composite supercell, that we call "supercell $C_1$", is then created.
4. A composite supercell $C_2$ is created by replacing the 1nn and the 2nn cations of the cation vacancy by means of the same procedure.
5. This procedure is applied up to the eighth coordination shell (8nn), i.e., up to the supercell $C_8$.

The supercell generation procedure described above is illustrated in Figure 4a in the case of 324-atom supercells and only for the first three shells. However, we have used 2592-atom supercells and applied the procedure up to the 8$^{th}$ shell. We call this procedure "implantation of SQS clusters in ordered supercells" and we apply it for each of the four ordered supercells considered and the ten SQS supercells generated.

Furthermore, we also generate composite supercells by replacing shell by shell the cations from the SQS supercells with the cations coming from the ordered supercells (see Figure 4b). This procedure is called "implantation of ordered clusters in SQS supercells".

In conclusion, with given SQS and ordered supercells, we create a set of eight composite supercells generated using the "implantation of ordered clusters in SQS supercells" and another set of eight composite supercells using the "implantation of SQS clusters in ordered supercells".

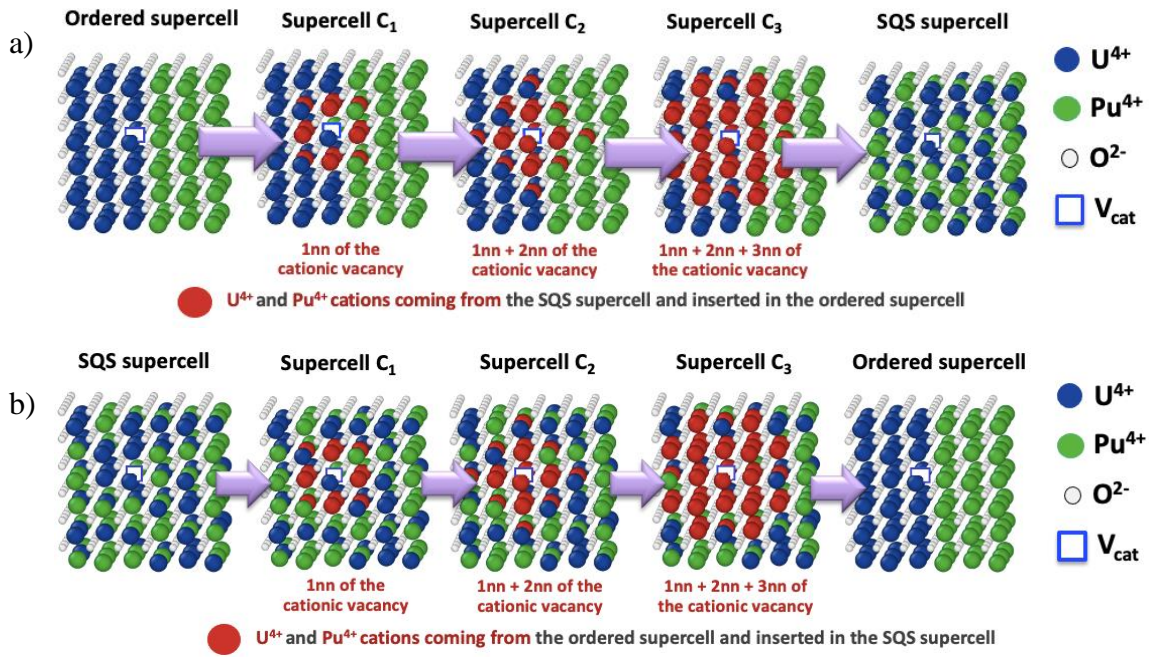

*Figure 4: Schematic representation of the procedure used to generate composite supercells from a SQS supercell and the 1/2-unmixed supercell in the case (a) "ordered in SQS generation" and (b) "SQS in ordered generation. Even though the procedure is here illustrated in the case of 324-atom supercells, calculations have actually been performed on supercells containing 2592 atoms.*

The composite supercells can be characterized by the radius of the shell up to which the cations from a supercell A are removed and substituted by the cations of a supercell B. The labels A and B refers to a SQS supercell or an ordered supercell, depending on the generation procedure considered. In Table 6 in Appendix C are summarized the radius of the first eight coordination shells on the *fcc* sublattice, and the number of atoms in each shell.

Let us note that, on each shell, the Pu atomic concentration is not equal to the nominal Pu atomic concentration of the whole supercell. Therefore, the nominal chemical composition varies slightly between each composite supercell. However, as one can see on Figure 16 in the Appendix C, the deviations do not exceed 1.5 % from the targeted chemical composition (50 %) imposed in the SQS and ordered supercells.

### III.2. Determination of the local environment cutoff distance.

In the present section, we present the results of our BSD formation energy calculations computed in the supercells generated according to the procedure described in the previous section. From this, we estimate the distance $R_D$ around a BSD from which the cationic chemical environment does not have a significant effect on its formation energy. Here, we show only the results obtained in the case of BSD formation energies calculated in the composite supercells generated from the ordered supercell called "plane 1" (see Figure 3a) and the ten SQS supercells. All other cases described in the previous section lead to the same conclusions. In Figure 5a and Figure 5b are shown the results for BSD3. The results for BSD1 and BSD2 are displayed in the Appendix C in Figure 17 and Figure 18, respectively.

We show in Figure 5a the results obtained in the case of composite supercells generated using the procedure called "implantation of ordered clusters in SQS supercells" (see Figure 4b). The horizontal black line refers to the BSD3 formation energy calculated in the ordered supercell plane 1. The results at $r = 0$ Å represent the formation energy in the SQS supercells, while those at increasing radii correspond to the formation energy calculated in the composite supercells (CS). We recall that in each of these CS, the cations around the defect in the SQS supercells are replaced by the ones of the ordered supercell up to a specific coordination shell, i.e., up to a specific distance which corresponds to the radius of the local environment. Thus, the BSD3 formation energy in the CS are shown as a function of the associated radius, which is unique for each CS. For instance, the composite supercells $C_1$ are associated to a radius equal to 3.84 Å (which is the radius of the first cation coordination shell). In Figure 5a, we count ten datasets: each of them refers to the CS generated by replacing the atoms of each SQS supercell by those of the ordered supercell plane 1.

It can be seen in Figure 5a that the BSD3 formation energy calculated in the ten different SQS supercells ranges between 4.56 eV and 5.01 eV, i.e., in a 0.45 eV interval. This value accounts for a minimal effect of cation disorder on BSD3 formation energy in $(U,Pu)O_2$. Concerning the BSD3 formation energy in CS showed in Figure 5a, we observe that when the radius of the local ordered environment increases, the formation energy converges to the value calculated in the ordered supercell represented by the black line (4.88 eV). This observation confirms our initial intuition that the formation energy is mainly governed by its close chemical environment. The values of the BSD3 formation energy computed in the supercells $C_3$ to $C_8$ show an almost constant deviation by a maximum of 0.05 eV from the BSD3 formation energy calculated in the ordered supercell plane 1. The small deviations that remain are due to the slight variations of chemical composition in the CS (see Figure 16 in Appendix C) and to the effect of chemical environment of the upper shells. Therefore, the chemical environment beyond the third nearest neighbors (3nn) of the defect does not significantly affect the BSD3 formation energy. Similar results are obtained in the cases of BSD1 and BSD2, as shown in Figure 17 and Figure 18 in Appendix C.

Figure 5b shows the data obtained in the case of the CS generated by "implanting SQS clusters in ordered supercells". The BSD3 formation energies calculated in the ten SQS supercells are shown with dashed lines with different colors. As in Figure 5a, the results associated to the CS are shown with dots linked by solid lines. We see that for each set of CS (each set associated to a specific SQS supercell), a difference of less than 0.05 eV is observed between the formation energy calculated in the composite supercells from $C_3$ to $C_8$ and the one obtained in their associated SQS supercell. The origins of the small deviations that remain are the same as those discussed in the previous paragraph. This result agrees with our previous observations from Figure 5a: only the cations up to the 3nn around a BSD have a significant influence on its formation energy. The same trends are found in the cases of BSD1 and BSD2 in Figure 17 and Figure 18 shown in Appendix C.

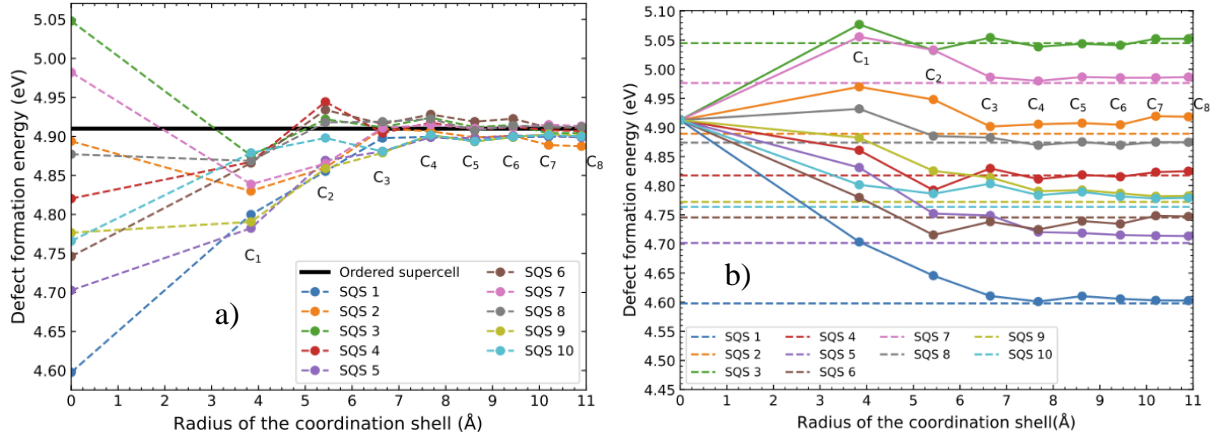

*Figure 5: BSD3 formation energy calculated in composite supercells generated by the implanting a) "ordered clusters in SQS supercells" and b) "SQS clusters in the ordered supercell plane 1". In a) the black line refers to the BSD3 formation energy calculated in the ordered supercell. In b) the dashed lines refer to the BSD3 formation energy in each of the ten SQS supercells.*

To sum up, our study allows us to conclude that the cationic chemical disorder beyond the 3nn cations around a BSD does not significantly modify its formation energy. In the case of $U_{0.5}Pu_{0.5}O_2$, the radius of the third coordination shell is equal to 6.65 Å and the radius of the fourth one is 7.68 Å. Thus, we can state that above a cutoff distance of $R_D \simeq 7.0$ Å the cationic chemical environment does not modify significantly the BSD formation energies in $(U,Pu)O_2$ (the energy difference is less than 0.05 eV). Therefore, the effect of cationic chemical disorder on this BSD property can be determined by considering cation configurations that include the 1nn-, 2nn-, and 3nn- cations.

## IV. Systematic evaluation of the impact of the local environment composition

### IV.1. Principle of the systematic study

In the present section, we aim at a systematic exploration of the cation configurations in the local atomic environment of a BSD, in order to study the influence of chemical disorder on the BSD formation energy. To achieve this goal, according to the findings of the previous section, we should consider all the possible distributions of U and Pu atoms in the cation sites of the first three coordination shells around the BSD cation vacancy. However, considering three neighboring shells leads to an unmanageable amount of configurations. If $k$ is the number of sites contained in the $n$-th shell, there are $2^k$ possible ways to distribute the U and Pu cations, i.e., $2^k$ possible cation configurations. The number of configurations to explore is extremely large, and clearly out of reach for DFT+$U$ calculations (*e.g.*, of the order of $10^{12}$ for $k = 3$). For this reason, we rely instead on empirical interatomic potentials calculations, more specifically the CRG potential [20] as in the previous section. However, even with EIP, the number of configurations to consider for the third coordination shell and beyond is too large. Therefore, we have performed our systematic study to the 1nn and 2nn nearest neighbor shells of the BSD cation vacancy. According to the analysis of Section III, we show that a restriction of BSD's neighbors to 1nn produces a discrepancy of 0.45 eV, which is much greater than the effect of the upper shells (*e.g.*, 0.12 eV for the 2nn) (resp. 0.12 eV).

As a first step, we generate reference supercells containing a BSD, namely 2592-atom supercells of $UO_2$, $PuO_2$, and $(U,Pu)O_2$ with three Pu nominal concentrations $y$ ($y$ = 0.25, 0.50, 0.75). Then, we introduce a systematic variation of the cation configurations on the 1nn and

2nn shell around the BSD in each reference cell. We first focus on the 1nn shell and try out all possible U/Pu combinations on the 12 first-nearest cations around the cation vacancy. This leads to 4096 ($2^{12}$) distinct cation configurations included in each reference supercell. In these supercells, we call "outer region" the chemical environment beyond the 1nn cation shell. However, if the cation sublattice beyond the 1nn shell is only filled with U or Pu (i.e., if the outer region beyond the first shell is pure oxide $UO_2$ or $PuO_2$, which are ordered distributions), some of the 4096 configurations are equivalent thanks to symmetries, which leads to a reduced number of configurations to explore. In this specific case, the reduced number of configurations to explore is 414 configurations around a BSD3, 1104 around a BSD2, and 1152 around a BSD1, as shown in Table 1.

Then, we extend the exploration to the 2nn shell following the same methodology. However, in this case we consider the BSD3 only for computational reasons. There are indeed $2^{18} = 262144$ possible configurations to explore. In supercells with a pure $UO_2$ and $PuO_2$ beyond the second coordination shell, we consider all possible U/Pu configurations on the 18 sites in the first two shells around the defect, which amounts to 22668 possible configurations when symmetries are considered (see Table 1).

Table 1: Amount of possible cation configurations up to the n-th (n=1,2) shell of the cation vacancy of the BSD by taking symmetries into account, i.e., whether the outer region beyond the n-th cation shell is a $UO_2$ or a $PuO_2$ environment.

| Type of Bound Schottky Defect | BSD1 | BSD2 | BSD3 |
|---|---|---|---|
| Number of nearest-neighbor configurations ($n = 1$) | 1152 | 1104 | 414 |
| Number of second-nearest-neighbor configurations ($n = 2$) | 68352 | 66816 | 22668 |

### IV.2. Results of the systematic study by means of interatomic-potential calculations

#### IV.2.1. Effect of cationic chemical disorder on the first coordination shell

In this section, we discuss the BSD formation energies computed with the CRG potential by varying the U/Pu composition on the first cation coordination shell of the cation vacancy. Figure 6 compares the BSD1, BSD2 and BSD3 formation energies for two different atomic environments in the outer region: pure $UO_2$ (Figure 6a) and a MOX with 25% Pu (Figure 6b). We show in Appendix D the results for the cases where the environment in the outer region (beyond the 1nn shell) contains 50%, 75%, and 100 % Pu (see Figure 19). In Figure 6b, the results at x-coordinate associated with zero Pu first nearest neighbors correspond to the BSD formation energies in $UO_2$, 6.17 eV, 5.27 eV and 5.04 eV for BSD1, BSD2, and BSD3 respectively. These values are in agreement with the studies of Balboa *et al.* [19] and Cooper *et al.* [20] who calculated them with the same empirical interatomic potential. We recall that we focus on the analysis of the BSD-U. However, we show in Appendix D the data obtained in the case of the BSD-Pu (see Figure 20). The results obtained in the case of a $PuO_2$ environment in outer region (beyond the 1nn cation shell) and twelve Pu on the cation vacancy 1nn sites correspond to the BSD formation energy in $PuO_2$ in Figure 20. We obtain 5.97 eV for BSD1,

5.05 eV for BSD2 and 4.81 eV for BSD3, which agree with the CRG potential calculations of Wang *et al.* [24].

Furthermore, Balboa *et al.* [19] obtained the following BSD formation energies in $U_{0.75}Pu_{0.25}O_2$: 6.5 ± 0.6 eV, 5.5 ± 0.6 eV and 5.4 ± 0.7 eV for BSD1, BSD2, and BSD3 respectively. The results of Balboa *et al.* are slightly larger than the values in our dataset shown in Figure 6b, for all compositions of the first cation shell. This mismatch can be explained by the fact that the authors of that work used a different energy reference (Equation 2.2 in place of Equation 2.1) for their calculations. On the contrary, our BSD formation energies in $U_{0.75}Pu_{0.25}O_2$ are larger than those computed by Cheik Njifon [15] using DFT+$U$, due to the different energy reference taken into account for their calculations.

Let us define as $\Delta E_F(BSDi) = E_{F,MAX} - E_{F,MIN}$ (i=1,2,3) the energy difference between the maximum $E_{F,MAX}$ and the minimum $E_{F,MIN}$ BSDi formation energies observed in Figure 6 (and in Figure 19 in Appendix D). This energy interval can be interpreted as the maximum effect of cationic disorder in the first coordination shell on the defect formation energy. Table 2 gathers the calculated values of $\Delta E_F$, and shows that this property depends on the type of BSD and on the nominal plutonium content in the supercell: namely, $\Delta E_F$ tends to increase with the nominal plutonium content. Moreover, $\Delta E_F$ is smaller in the case of BSD2 than in that of BSD3 and BSD1: it can vary indeed from 0.83 eV to 0.87 eV for BSD1, 0.84 eV to 0.89 eV for BSD3, and 0.70 eV to 0.74 eV for BSD2, depending on the chemical composition of the outer region beyond the 1nn shell.

Furthermore, it can be observed in Figure 6 (and in Figure 19 and Figure 20) that all the values obtained in the case of BSD1 are larger than those obtained for BSD2 and BSD3, whereas the ranges of possible values for BSD2 and BSD3 partially overlap. This means that a BSD2 can have a lower formation energy than BSD3 depending on the configuration of the first coordination shell, and vice versa. However, for a given local configuration, BSD3 always has a lower formation energy than BSD2. Let us notice that in this case, we call "outer region" the chemical environment beyond the 2nn cation shell.

*Table 2: Calculated values of the energy interval $\Delta E_F = E_{F,MAX} - E_{F,MIN}$, with $E_{F,MAX}$ ($E_{F,MIN}$) the maximum (minimum) BSD formation energy, obtained with the CRG potential, for different composition of the environment beyond the first coordination shell of the defect.*

|  | $\Delta E_F$ for BSD1 (eV) | $\Delta E_F$ for BSD2 (eV) | $\Delta E_F$ for BSD3 (eV) |
|---|---|---|---|
| 0 % Pu | 0.83 | 0.70 | 0.84 |
| 25 % Pu | 0.84 | 0.70 | 0.85 |
| 50 % Pu | 0.84 | 0.71 | 0.86 |
| 75 % Pu | 0.86 | 0.72 | 0.87 |
| 100 % Pu | 0.87 | 0.74 | 0.89 |

In order to get further insight into the effect of chemical disorder, the results shown in Figure 6 are analyzed further in Figure 7, Figure 8, and Figure 9 for BSD1, BSD2, and BSD3 respectively, with a color coding that indicates the number of Pu atoms that are first nearest neighbors of an oxygen vacancy (1nn-$V_O$). These figures demonstrate two physical effects due to the presence of Pu in the first cation coordination shell.

Let us consider the case of BSD3 with 0 % Pu in the outer region (in Figure 9a). The data circled in red show that, as we increase the number of Pu atoms next to an oxygen vacancy, the formation energy decreases, until a minimum formation energy is reached when all the neighbors of the oxygen vacancies are all filled with Pu (see Figure 2). The configuration leading to this minimum energy is shown on Figure 10f. Then, the data circled in green show that increasing the number of Pu atoms on the remaining sites (that are necessarily not direct neighbors of the oxygen vacancies) makes the formation energy increase linearly, until the whole 1nn shell around the cation vacancy is filled with Pu atoms. Similar effects are seen in the case of BSD1 and BSD2 in Figure 7 and Figure 8, respectively. For BSD1, however, there are only 5 sites that are nearest neighbors of the oxygen vacancies instead of 6 (see Figure 2). The 1nn cation configurations that are associated to the minimum BSD1 and BSD2 formation energies are shown in Figure 10b and Figure 10d respectively. Thanks to this analysis, we can deduce that the maximum BSD formation energy is obtained when all the oxygen vacancy nearest neighbors are occupied by uranium cations: the corresponding cation configurations for BSD1, BSD2 and BSD3 are shown in Figure 10a, Figure 10c, and Figure 10e respectively.

To confirm the CRG potential results shown in the present section, we have performed DFT+$U$ calculations on a small selected subset of 1nn configurations circled in Figure 6a. This study is presented in Section V.

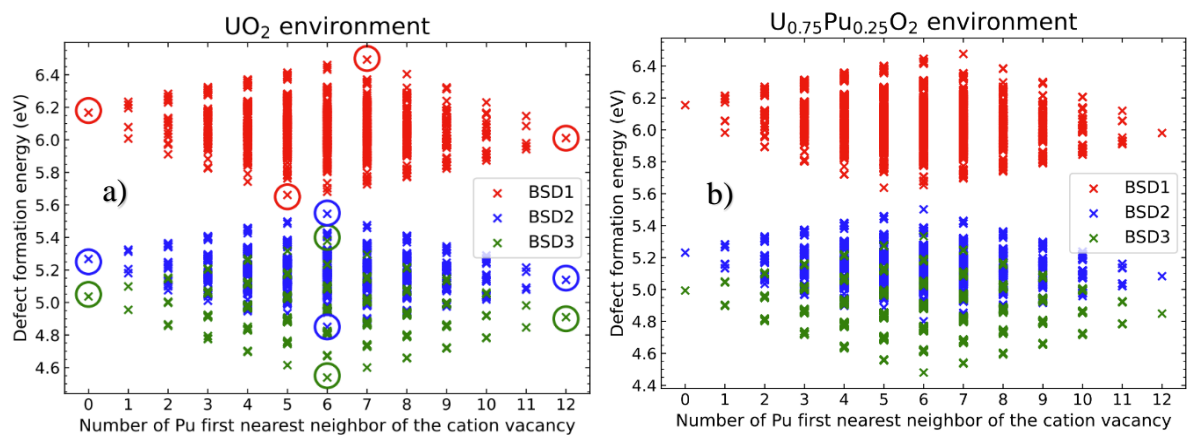

*Figure 6: BSD1 (red), BSD2 (blue) and BSD3 (green) formation energies calculated using the CRG potential with (a) only U atoms and (b) 25% Pu on the cation sublattice beyond the first coordination shell. The circled data are the selected configuration for DFT+U calculations in section V.*

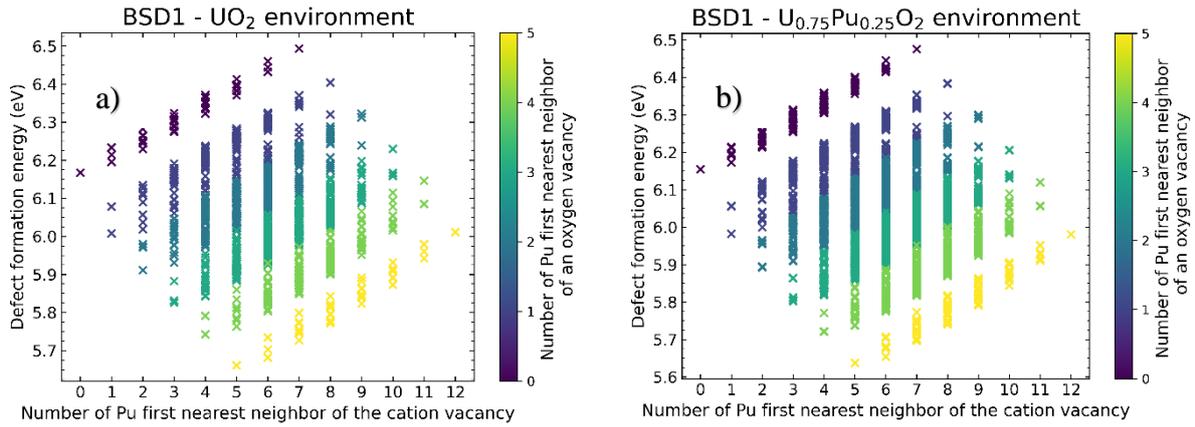

*Figure 7 : BSD1 formation energy as a function of the first nearest neighbor (1nn) cation configurations in (U,Pu)O$_2$ with (a) only U atoms on the cation sublattice beyond the first coordination shell and (b) 25% Pu. The color code represents the number of Pu atoms that are 1nn of the oxygen vacancies.*

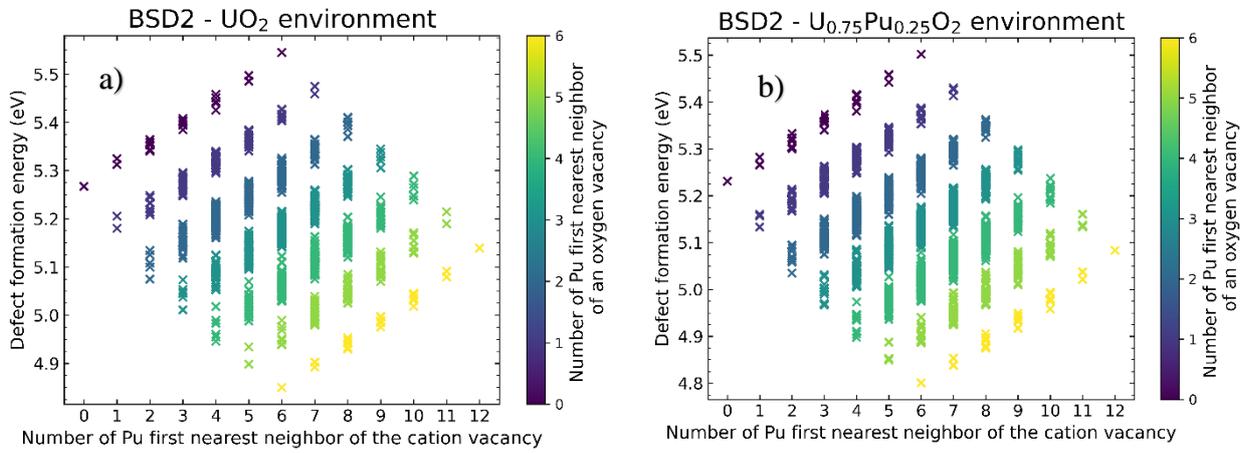

*Figure 8 : BSD2 formation energy as a function of the nearest neighbor (1nn) cation configurations in (U,Pu)O$_2$ with (a) only U atoms on the cation sublattice beyond the first coordination shell and (b) 25% Pu. The color code represents the number of Pu atoms that are 1nn of oxygen vacancies.*

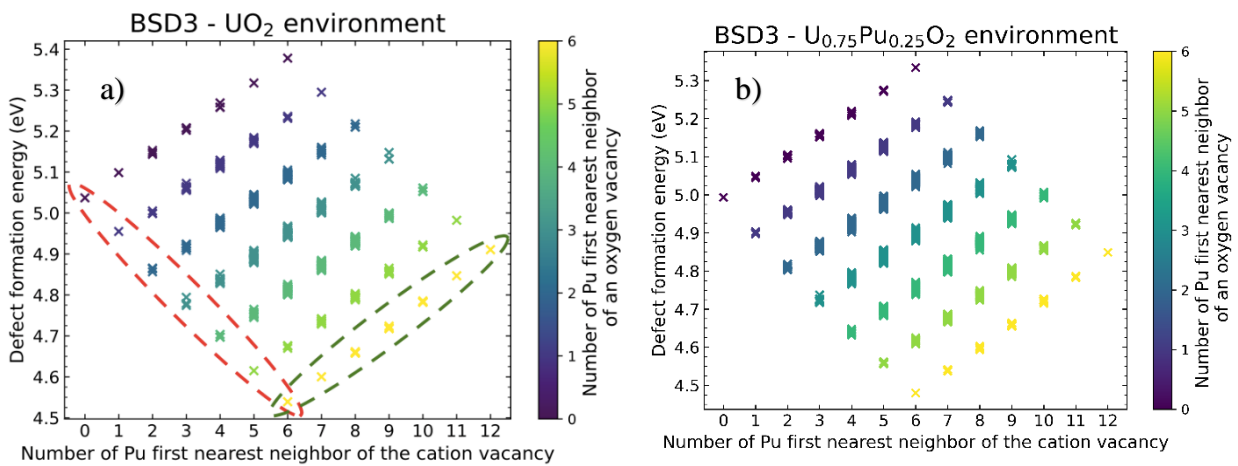

*Figure 9 : BSD3 formation energy as a function of the first nearest neighbor (1nn) cation configurations in (U,Pu)O$_2$ with (a) only U atoms on the cation sublattice beyond the first coordination shell and (b) 25% Pu. The color code represents the number of Pu atoms that are 1nn of an oxygen vacancy. Dashed circles gather data obtained when all the Pu atoms of the first cation shell are 1nn of an oxygen vacancy.*

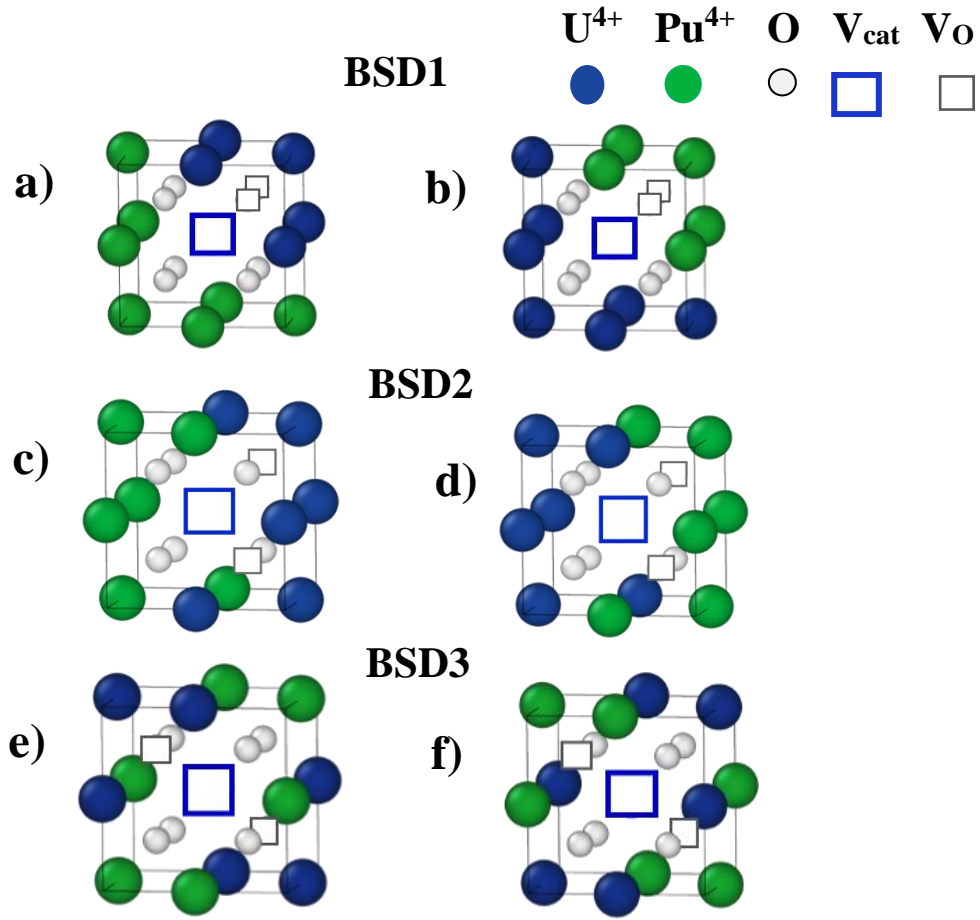

*Figure 10: First nearest-neighbor cation configurations associated with the maximum and the minimum BSD formation energies. (a), (c), (e) are the configurations related to the maximum formation energies for BSD1, BSD2 and BSD3 respectively. (b), (d) and (f) are the configurations related to the minimum ones.*

### IV.2.2. Effect of nominal plutonium concentration in the supercell

The effect of nominal plutonium concentration in the supercell on the BSD1, BSD2, and BSD3 formation energies is shown in Figure 11a, Figure 11b, and Figure 11c, respectively. We observe that for a given cation nearest neighbor configuration:

- The nominal Pu content modifies the formation energy by 0.09 eV at most in the cases of BSD2 and BSD3, and by 0.05 eV in the case of BSD1.

- The BSD1 and BSD3 formation energies with a pure $PuO_2$ and $UO_2$ environment in the outer region are always the lowest and the largest, respectively.

- The BSD1, BSD2 and BSD3 formation energies obtained for intermediate $(U,Pu)O_2$ compositions do not show any clear trend with respect to the nominal Pu content, but for BSD1 and BSD3, they always remain between the values corresponding to the $UO_2$ or $PuO_2$ environments beyond the 1nn cation shell.

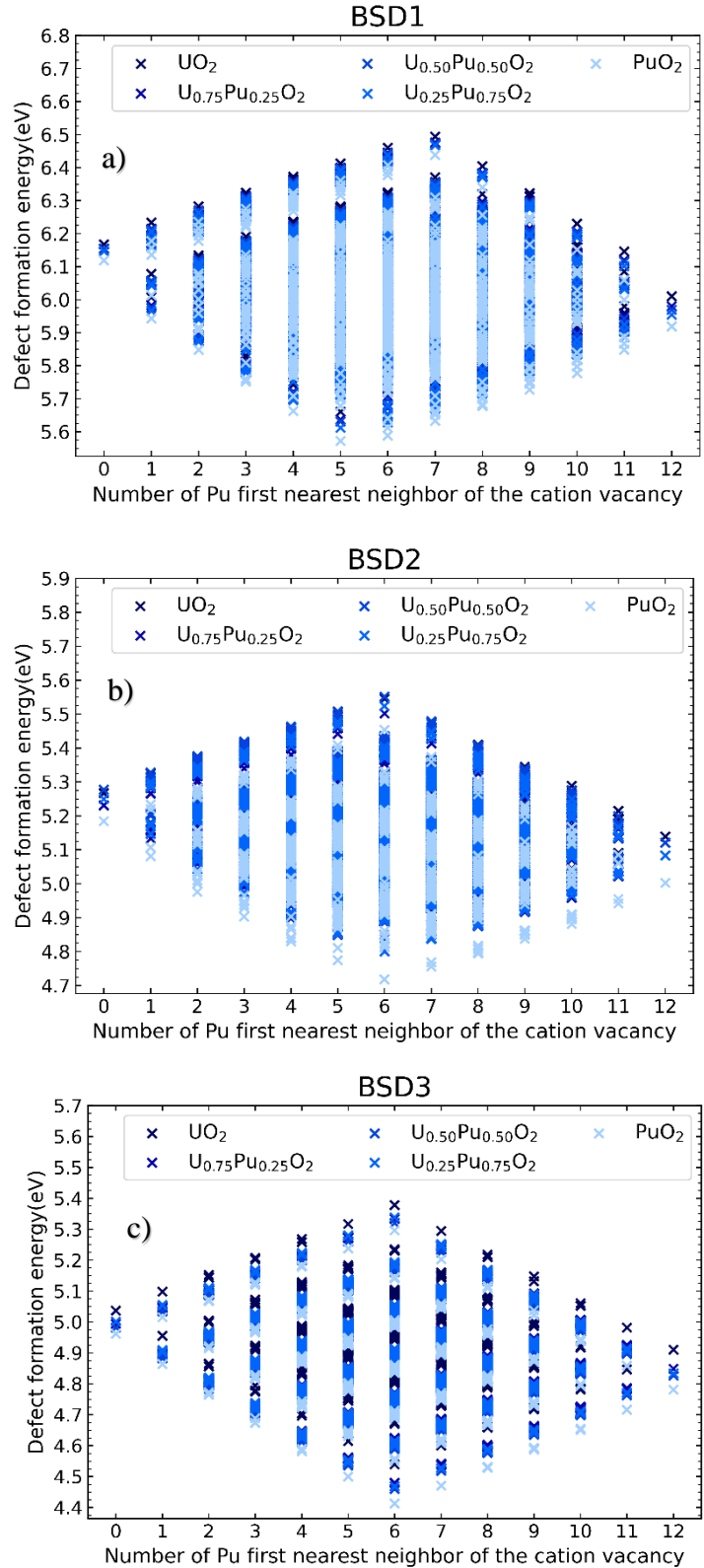

*Figure 11: Effect of nominal plutonium content on the a) BSD1, b) BSD2, c) BSD3 formation energies, for different chemical compositions beyond the first coordination shell of cation vacancy.*

In conclusion, the nominal Pu atomic concentration does not modify the value of $\Delta E_F$, the linear variation of $E_F$ with the number of 1nn Pu atoms, nor the change of sign of this variation when the number of 1nn Pu atoms reaches 6 (i.e., the shape of the diamond-shaped graphs).

### IV.2.3. Statistical correlations

In order to better understand the interplay between U/Pu and vacancies in the BSD formation energy, we analyze the statistical correlation between the number of Pu in the near environment of the BSD and their formation energies. Let us define $n(Pu\text{-}V_O)$ the number of Pu 1nn of an oxygen vacancy $V_O$ and $n(Pu\text{-}V_{cat})$ the number of Pu 1nn of the cation vacancy $V_{cat}$. An efficient way to analyze the linear correlation between the BSD formation energies $E_F$ in $(U,Pu)O_2$ and the quantities $n(Pu\text{-}V_O)$ or $n(Pu\text{-}V_{cat})$ is to compute the Pearson's correlation coefficient $r$. For two random variables $X$ and $Y$, $r$ (or $r[X, Y]$) is defined as follows:

$$r[X,Y] = \frac{\text{cov}(X,Y)}{\sigma_X \sigma_Y}, \qquad (4.1)$$

with $\sigma_X$ and $\sigma_Y$ the standard deviation of the random variables $X$ and $Y$, and $\text{cov}(X,Y)$ their covariance. We compute $r[X, Y]$) with $X = E_F$ and $Y = \{n(Pu\text{-}V_O), n(Pu\text{-}V_{cat})\}$ for the BSD formation energy data obtained for each chemical composition of the outer region. The results are summarized in Table 3. The calculated values are comprised between 0 and 1: a null Pearson's coefficient means that the two random variables are uncorrelated. On the contrary, if $r = \pm 1$, a perfect linear correlation exists. $r[E_F, n(Pu\text{-}V_O)]$ values range between -0.843 and -0.805 for all BSD and all combined compositions of the outer region. These values are much larger than $r[E_F, n(Pu\text{-}V_{cat})]$ ones, which range between -0.182 and -0.152. This means that the values of BSD formation energies $E_F$ are much more correlated to the amount of Pu 1nn of $V_O$ than the amount of Pu 1nn of $V_{cat}$. This is consistent with the linear behavior of $E_F$ with respect to $n(Pu\text{-}V_O)$ described in the previous section, whereas we can see that $E_F$ can decrease or increase as a function of $n(Pu\text{-}V_{cat})$. Furthermore, our calculated $r$ values are negative. This means that our BSD formation energy data tend to decrease by increasing $n(Pu\text{-}V_O)$ and $n(Pu\text{-}V_{cat})$, which agrees with our previous observations. Our $r$ values in Table 3 do not depend significantly on the nominal Pu concentration. Some differences are observed for the cases of 0 % and 100 % Pu environment in the outer region, because there are fewer data in these specific cases. Therefore, the results detailed in this paragraph could explain the low impact of Pu nominal concentration on the BSD formation energy, as the latter property is mainly impacted by the local interactions of the oxygen vacancies with their nearest surroundings cations.

*Table 3 : Pearson's coefficient measuring the linear correlation between the number of Pu 1nn of cation n(Pu-V$_{cat}$) or oxygen vacancies n(Pu-V$_O$) and the value of BSD formation energies E$_F$ calculated with the CRG potential in the framework of the systematic study.*

|  | $E_F$(BSD1) | | $E_F$(BSD2) | | $E_F$(BSD3) | |
|---|---|---|---|---|---|---|
|  | $n(Pu\text{-}V_{cat})$ | $n(Pu\text{-}V_O)$ | $n(Pu\text{-}V_{cat})$ | $n(Pu\text{-}V_O)$ | $n(Pu\text{-}V_{cat})$ | $n(Pu\text{-}V_O)$ |
| 0 % Pu | -0.168 | -0.834 | -0.177 | -0.811 | -0.152 | -0.805 |
| 25 % Pu | -0.186 | -0.842 | -0.201 | -0.823 | -0.165 | -0.813 |
| 50 % Pu | -0.194 | -0.843 | -0.210 | -0.831 | -0.175 | -0.819 |
| 75 % Pu | -0.199 | -0.847 | -0.218 | -0.833 | -0.187 | -0.825 |
| 100 % | -0.209 | -0.857 | -0.236 | -0.845 | -0.204 | -0.835 |

### IV.2.4. Effect of cationic chemical disorder up to the second coordination shell

In the present section, we study the effect of cationic chemical disorder on the second nearest neighbor shell. As mentioned in Section IV.1, the cost of computing the formation energy in all possible U/Pu configurations on the first two coordination shells (exactly $2^{18} = 262144$) is too large even for interatomic-potential calculations. We showed previously (see Table 1) that symmetries induce a drastic reduction of the number of cation configurations that we need to explore in case the composition beyond the 2nn cation shell is pure $UO_2$ or pure $PuO_2$. In this case, a systematic exploration entails 22668 configurations for BSD3, 66816 for BSD2, and 68352 for BSD1. Therefore, we limited our study to the cases of BSD3 with an outer region beyond the 2nn cation shell filled with pure $UO_2$ and $PuO_2$ environments. The results are shown in Figure 12a (Figure 12b), namely the BSD3 formation energies as functions of the number of Pu atoms in the 1nn shell for a $UO_2$ ($PuO_2$) outer region. The color code refers to the number of 2nn Pu atoms of the cation vacancy.

Figure 12 shows that the energy interval $\Delta E_F$ separating the maximum and minimum formation energies is larger than the one observed in the previous section (see Table 2), with an increase from 0.84 to 0.97 eV in the $UO_2$ environment, and from 0.89 to 0.96 eV in $PuO_2$. This increase is larger than the effect of nominal plutonium concentration observed in the previous section (Figure 11c). We can also observe that the formation energy decreases when the number of Pu atoms on the 2nn shell increases.

In Figure 13, we compare the results obtained in the case of the BSD3-U with the ones related to the BSD3-Pu. The data are shown as functions of the total number of Pu atoms in the first and second coordination shells. The data for the BSD3-Pu and the BSD3-U follow a similar trend with respect to the number of Pu in the first and second coordination shells: the difference lies in a slight shift (lower than 0.05 eV) due to the different energy references considered in Eq. 2.2.

Based on the observations made in the present section, in Section IV.2.1 and in Section IV.2.2, we propose an analytical law for the formation energy of a BSD3 that depends on the composition of its local environment as follows:

$$E_F(\text{BSD3}) = A + A_0\ N_{1nn}^O + A_1\ N_{1nn}^{cat} + A_{12}\ N_{1nn}^{cat\ 2} + A_2\ N_{2nn}^{cat} + A_{22}\ N_{2nn}^{cat\ 2}, \quad (4.2)$$

with $N_{1nn}^O$, $N_{1nn}^{cat}$ and $N_{2nn}^{cat}$ respectively the number of Pu 1nn of an oxygen vacancy, 1nn and 2nn of the cation vacancy. We include the linear behavior of $E_F$ with respect to $N_{1nn}^O$, in accordance with our observations in sections IV.2.1 and IV.2.2, by adding a linear term in Eq. (4.2). We add quadratic terms $N_{1nn}^{cat\ 2}$ and $N_{2nn}^{cat\ 2}$ to take into account the non-linear effects of Pu addition in the 1nn and 2nn shells of the cation vacancy respectively. Coherently with the observation in Section IV.2.2 that the defect formation energy is not significantly affected by the composition of the outer region, we do not include the effect of Pu nominal concentration. It was indeed shown in the previous section that its effect is lower than the influence of cationic disorder, and that no clear trend between $E_F$ and the nominal chemical composition exists (see Section IV.2.2). By fitting the analytical expression on the data shown in Figure 13, we obtain the values of the parameters summarized in Table 4. The difference between the law and the calculated data is always less than 1%. Thus, we observe a good agreement between the data of the law and our CRG potential results. This agreement is confirmed by the root-mean-square

(RMS) value of the law (1.5·10⁻⁴). Our law summarizes in a relatively simple way a large number of calculations, and can be used as a valid input interaction model in high-scale simulations.

In conclusion, we find that the formation energy of a BSD defect vary by 0.97 eV depending on the composition of its local environment, essentially the specific positions of Pu atoms with respect to the oxygen vacancies. This variation is well represented by the analytical law (Eq. *4.2*), valid in the case of BSD3. The 0.97 eV interval is an unexpected result because the gap between the values of BSD3 formation energies in $UO_2$ (5.04 eV) and $PuO_2$ (4.81 eV) is only 0.23 eV according to the CRG potential. It is actually explained in Ref [25] that the point defect formation energies also depend on the alloy ordering. For instance, in an alloy AB with a clustering tendency, the minor element tends to segregate on the neighboring lattice sites of the point defect to reduce the number of non-favorable A-B bonds. This explains the parabolic-like variation of the BSD formation energy with the number of 1nn Pu atoms. The positive curvature of the formation energy convex hull highlights a clustering tendency between Pu and U. The width of the interval will need to be confirmed by further studies exploring the effect of the 3nn shell, which should have a non-negligible impact according to the results shown in Section III.2. Furthermore, it is important to note that the configurations shown in all diamond-shaped graphs are not equiprobable, and this can have an impact on the actual width of the interval. First, from a purely configurational entropy standpoint, most configurations are concentrated toward the center of the diamond, rather than on the edges. In other words, the extreme $E_F$ values belong to few specific configurations and do not represent most of the possible values. Moreover, the distribution of local chemical compositions around a defect is not uniform, but depends on the nominal concentration. For instance, in (U,Pu)$O_2$ with 50% Pu, local environments with 100% Pu atoms are much rarer than those with an equal mix of U and Pu cations. Finally, we neglect in the present study the possible effect of local order, i.e., any possible deviation from the ideal solution behavior (induced or not by the presence of the defect) that can make some configurations more probable than others. Additional studies are planned to analyze possible local-order effects, and to assess the effect of the 3nn shell, using for the latter machine-learning methods suited to explore the enormous configuration space.

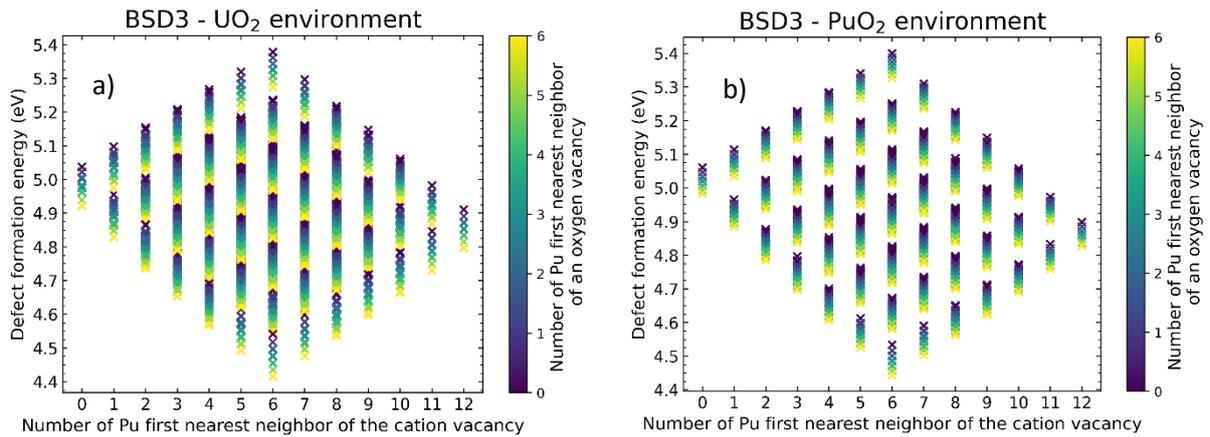

*Figure 12: BSD3 formation energies for all possible cation configurations on the first and second coordination shells of the cation vacancy, when the cations in the environment beyond the second coordination shell are only (a) U cations or (b) Pu cations. The colors refer to the number of Pu on the second coordination shell.*

Table 4: Values of the parameters of the analytical model of BSD3 formation energy in (U,Pu)O$_2$.

| Parameters | A | A$_0$ | A$_1$ | A$_{12}$ | A$_2$ | A$_{22}$ |
|---|---|---|---|---|---|---|
| Values (eV) | 5.0339 | - 0.1437 | 5.1752 | 6.80·10$^{-4}$ | -1.76·10$^{-2}$ | 1.81·10$^{-4}$ |

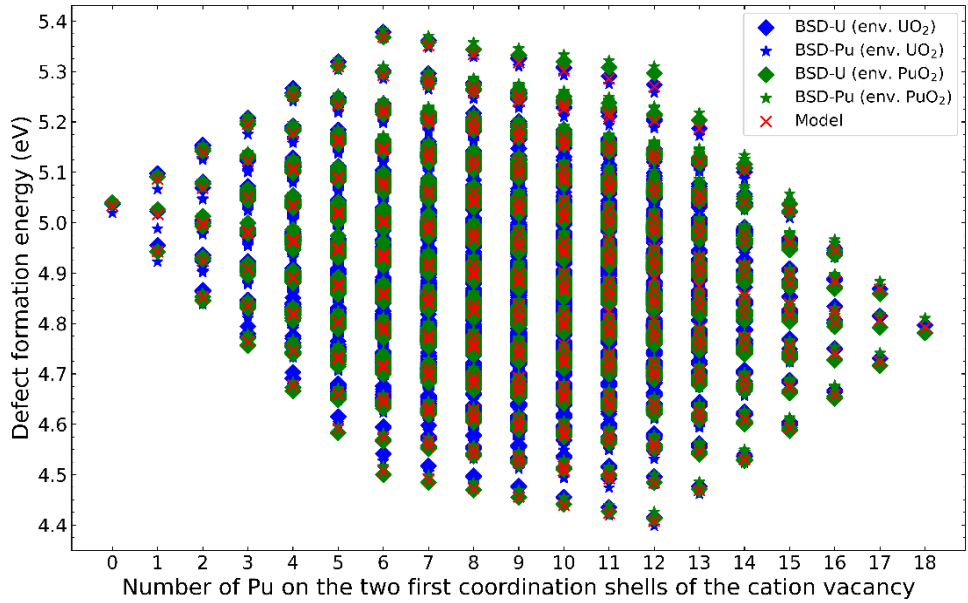

Figure 13: BSD3 formation energy for all possible cation configurations up to the second coordination shell of the cation vacancy, when the cations on the fcc sublattice beyond the second coordination shell are only U cations (blue markers) and only Pu cations (green markers). The red crosses are the results of the interaction model fitted on the calculated data.

## V.  Benchmark by DFT+$U$ calculations

In the present section, we detail the DFT+$U$ simulations performed to benchmark the results of the CRG potential. As electronic structure methods are computationally expensive, we limit our DFT+$U$ study to a small selected subset of 1nn configurations, shown in Figure 6a (Section IV.2.1). We chose the configurations that provide the maximum and minimum BSD formation energies, in the aim of confirming the energy interval ΔE$_F$ found with the CRG potential. We also calculate the formation energies for the configurations containing 12 Pu or 12 U in the first coordination shell. The DFT+$U$ simulations are run in two 324-atom supercells (the generation procedure is explained in Section V.1): the first case with only U cations beyond the first coordination shell, and the second case with only Pu. Then, considering configurations that contain 12 Pu or 12 U in the first coordination shell is equivalent to computing BSD formation energies in:

- pure UO$_2$.
- pure PuO$_2$.
- a shell of 12 Pu first nearest neighbors surrounded by only U cations.
- a shell of 12 U first nearest neighbors surrounded by only Pu cations.

## V.1. Computational details

The electronic structure calculations are performed with the Vienna Ab initio Simulation Package (VASP) code [26], [27] using the projector augmented wave [28], [29] (PAW) pseudopotential formalism. We use the vdW-optPBE exchange-correlation functional [30] in the framework of the generalized gradient approximation (GGA). A plane-wave basis set with a cutoff of 500 eV is chosen to describe the valence electronic states. The strong correlations among the $U^{4+}$ and $Pu^{4+}$ 5$f$ electrons are taken into account by adding an onsite Coulomb repulsion in the form of a Hubbard-type term ($+U$) in the Hamiltonian [16], using the rotationally invariant approach introduced by Liechtenstein *et al.* [31]. The $U$ and $J$ parameters for uranium cations are fixed to 4.50 eV and 0.54 eV respectively, in accordance with previous theoretical studies [32], [33]. For plutonium cations, $U$ is set to 4.00 eV and $J$ to 0.70 eV in agreement with Jomard *et al.* [34].

All calculations are performed in 3x3x3 supercells containing 324 atoms, which are built by including the selected 1nn cation configurations around a BSD in 324-atom pure oxide supercells. We consider in the present study the 1-$k$ antiferromagnetic (1$k$-AFM) order. The 3-$k$ AFM ground state for $UO_2$ is commonly accepted as the lowest energy state in the literature [35], [36]. Nonetheless, the 1$k$-AFM ground state is a reasonable approximation to model the magnetic order of uranium dioxide using DFT+$U$ calculations according to Dorado *et al.* [37]. Concerning $PuO_2$, non-magnetic ground states are determined by crystal field calculations and experimental studies [38]–[41]. Recent DFT+$U$ calculations found a 3-$k$ AFM ground state [42], but the 1-$k$ AFM order considered in a previous study [32] yields good results concerning structural and elastic properties compared to experimental data. We optimize the cell shape, cell volume, and atomic positions using the occupation matrix control scheme [43] for all systems, without symmetry constraints. Ionic forces are evaluated using the Hellman-Feynman theorem [44], and the Davidson-block iteration scheme [45] is used for the electronic minimization algorithm. The convergence criterion over self-consistent cycles is set to $10^{-7}$ eV per atom, and calculations are performed at the $\Gamma$ point of the Brillouin zone (single $k$-point).

## V.2. Results and discussion

In Figure 14, we compare the BSD formation energies calculated using DFT+$U$ (red markers) with those obtained with the CRG potential (blue markers) presented in the previous section. The square and diamond markers are the values calculated for the cation configurations with respectively 0 Pu and 12 Pu on the 1nn shell of the cation vacancy. The star and circle markers are the values calculated for the cation configurations associated with the maximum ($E_{F,MAX}$) and minimum ($E_{F,MIN}$) BSD formation energies according to the CRG potential. These targeted configurations correspond to those circled in Figure 6a, and some of them are illustrated in Figure 10.

The BSD formation energies in $UO_2$ (square markers in Figure 14a, Figure 14c, and Figure 14e) and $PuO_2$ (diamonds markers in Figure 14b, Figure 14d and Figure 14f) are compared with data available in the literature in Table 5. The formation energies calculated in $UO_2$ and $PuO_2$ using DFT+$U$ are lower than the values obtained with the CRG potential, except for BSD3 in $PuO_2$. Furthermore, according to DFT+$U$ calculations, the BSD formation energies in $UO_2$ are lower than those in $PuO_2$, as opposed to the results obtained with the CRG potential. Concerning $UO_2$, our calculations are close to the results obtained by Vathonne [12] who also used the vdW-optPBE functional but smaller supercells (96 atoms) than the ones used in the present study (324 atoms). However, unlike Vathonne we found that the BSD3 is more stable than BSD2 in

UO$_2$, and this result is confirmed by both DFT+$U$ and the CRG potential. This is due to the different supercell size, which has a non-negligible effect on defect formation energies in UO$_2$ according to Burr *et al.* [42]. Concerning PuO$_2$, the DFT+$U$ formation energies are rather close to those obtained with the CRG potential, although according to DFT+$U$ the BSD2 is more stable than BSD3, as opposed to what was predicted by CRG. Our DFT+$U$ results using 324-atom supercells are close to the DFT+$U$ of Cheik Njifon [15] performed using 96-atom supercells, but in these calculations, the BSD3 was more stable than BSD2. Hence, some discrepancies remain between the results available in the literature and ours. These discrepancies probably arise from the different supercell sizes, and, in the case of Cheik Njifon, from the different functional (PBE instead of vdW-optPBE).

The data on Figure 14a, Figure 14c, and Figure 14e (Figure 14b, Figure 14d, and Figure 14f) refer to the case of the BSD-U (BSD-Pu). Similar results are obtained for the BSD-Pu (BSD-U) with a slight shift lower than 0.05 eV. An equivalent shift was observed in our previous results with the CRG potential in Section IV.

In a UO$_2$ environment beyond the 1nn cation shell (Figure 14a, Figure 14c, and Figure 14e) we can observe a rather good qualitative agreement between the energy interval $\Delta E_F = E_{F,MAX} - E_{F,MIN}$ obtained using DFT+$U$ and the CRG potential in the cases of BSD1 and BSD3. Let us note that the data with 12 Pu in the 1nn shell in the case of BSD2 (Figure 14c) is missing because of a very slow convergence. We observe a reduction of the $\Delta E_F$ interval, namely 0.66, 0.32, and 0.71 eV (for BSD1, BSD2, and BSD3 respectively) with respect to 0.83, 0.70 and 0.84 eV (see Table 2). The ratios $\Delta E_F / E_{F,MAX}$ are indeed similar for BSD1 and BSD3 with DFT+$U$ and the CRG potential (0.16 in both cases for BSD3, and 0.13 vs 0.14 for BSD1). However, we observe a mismatch for the BSD formation energies when the 1nn shell is filled with 12 Pu: while the CRG potential predicts a lower formation energy with respect to the one in UO$_2$, this result is not confirmed by DFT+$U$. For this reason, the rhombus formed by the CRG potential data is not exactly reproduced by DFT+$U$, which means that the qualitative agreement between the two methods can be improved.

Concerning the BSD in a PuO$_2$ environment in the outer region (Figure 14b, Figure 14d, Figure 14f), we observe the same trends as in Figure 14a: we obtain with DFT+$U$ lower formation energies and lower values of $\Delta E_F$ with respect to CRG. Let us note that the red square markers for BSD1 and BSD2, as well as the diamond marker for BSD3 are missing because of a slow convergence. In Figure 14b, Figure 14d and Figure 14f, some DFT+$U$ results are in contradiction with the CRG predictions. For instance, the configurations giving the minimum and maximum BSD1 formation energy are swapped. Furthermore, in most cases the configurations that, according to CRG, should yield the maximum formation energy have actually a lower energy than other ones according to DFT+$U$. These discrepancies could have various origins. In order to investigate them in more detail, we show in Figure 15 the BSD formation energies calculated with the CRG potential versus the values obtained using DFT+$U$. Solid markers represent the data found in the case of a UO$_2$ environment in the outer region and empty markers refer to the case of a PuO$_2$ environment. The data in Figure 15 are split in two separated sets that we denote "set A" and "set B". Most of the values are contained in the set A and show a common linear trend drawn in gray in Figure 15. Therefore, it can be assumed that there is a linear correlation between the CRG potential values and DFT+$U$ data of the set A. This seems to suggest a rather good agreement between the two calculation methods for set A, as the discrepancies between the two are ascribed to a simple scaling factor. Let us note that we also find a scaling factor by calculating the reference energy E(XO$_2$)/N (see Eq. 2.2) is equal to -27.06 eV (-29.57 eV) for UO$_2$ (PuO$_2$) according to DFT+$U$, and to -40.68 eV (-41.94 eV) according to the CRG potential. The ratio between the two calculated reference energies is $E_{ref}(DFT) / E_{ref}(CRG) \approx 0.66$. Thus, we could expect a similar scaling factor among our calculated $E_F$ data. For the data contained in the set A, the ratio $E_F(DFT) / E_F(CRG)$ ranges between 0.68 and 0.88.

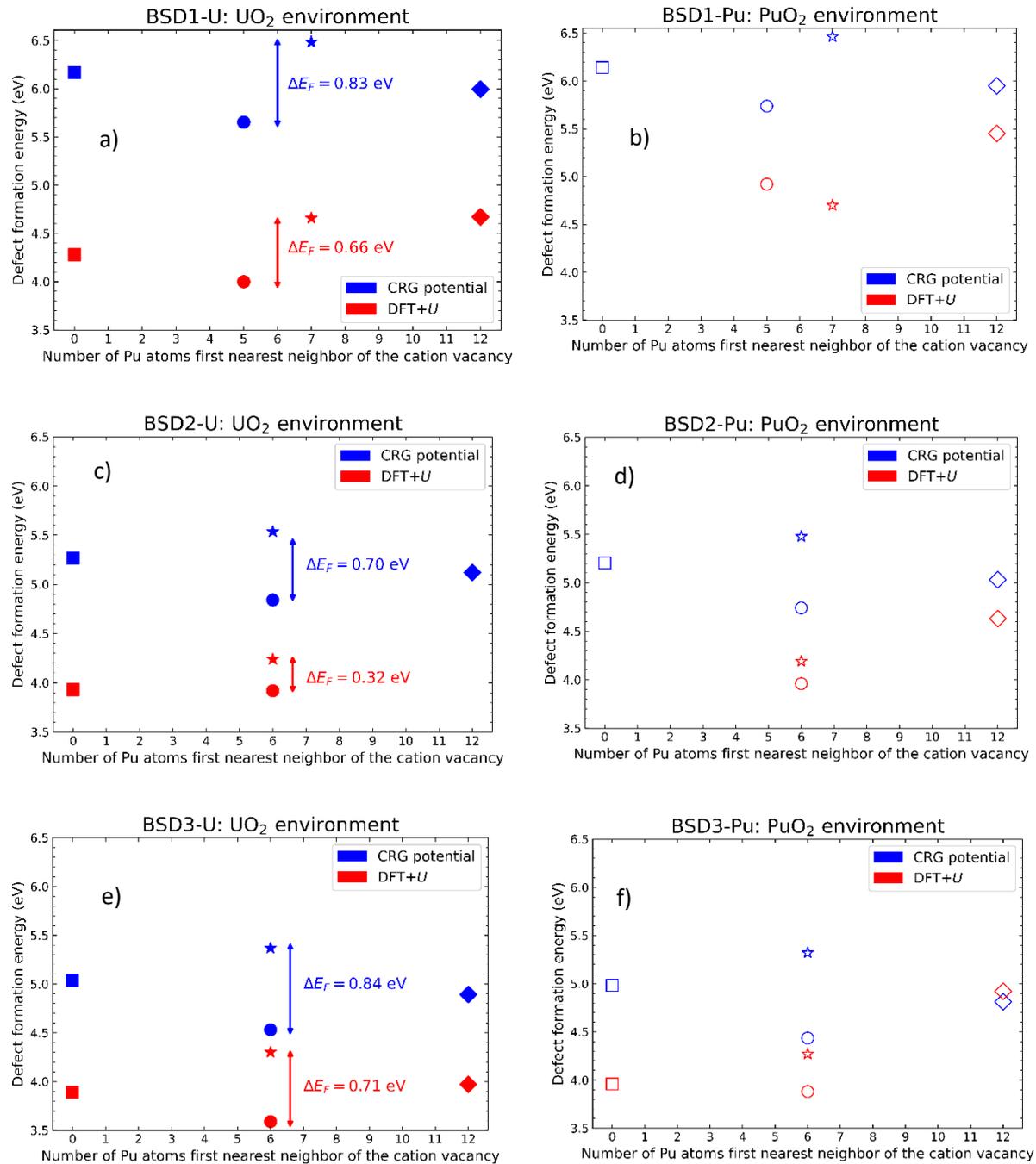

Figure 14: Comparison of the results obtained using DFT+U (red markers) and the CRG potential (blue markers). The labels $UO_2$- and $PuO_2$ environment refers to the nominal composition beyond the first coordination shell.

The set B contains only four data: the BSD1, BSD2, and BSD3 formation energy in $PuO_2$, as well as the BSD1 formation energy with a $PuO_2$ environment obtained for the configuration associated to the minimum BSD1 formation energy. Let us note that these four data correspond to those with the highest mismatch in Figure 14b, Figure 14d, Figure 14f as we previously discussed in this section. One can notice that the data deviating from the linear trend belong all to the case of a Pu-rich local environment. Even though a qualitative agreement is observed in most cases, the scaling factor introduces a systematic error in the point-defect properties predicted by the CRG, which should be investigated further. The comparison of our data with experimental values would be a reliable strategy to find out the origin of for these discrepancies. However, experimental data of BSD formation energy in $PuO_2$ are not available in the literature.

Table 5: Comparison of the BSD formation energies (in eV) in $UO_2$ and in $PuO_2$ using DFT+U and the CRG potential from this study and the literature. Between brackets are shown the values of $\Delta E_F = E_{F,MAX} - E_{F,MIN}$ (in eV) deduced for each BSD in the case of a $UO_2$ environment beyond the 1nn shell.

| Methods | DFT+$U$ | | DFT+$U$ (this study) | | CRG potential (this study) | |
|---|---|---|---|---|---|---|
| Number of atoms in the supercell | 96 | | 324 | | 2592 | |
| Pure oxides | $UO_2$ [12] | $PuO_2$ [15] | $UO_2$ | $PuO_2$ | $UO_2$ | $PuO_2$ |
| BSD1 | 4.01 | 5.57 | 4.24 (0.66) | 5.49 | 6.17 (0.83) | 5.95 |
| BSD2 | 3.33 | 4.85 | 3.93 (0.32) | 4.63 | 5.27 (0.70) | 5.04 |
| BSD3 | 3.43 | 4.32 | 3.89 (0.71) | 4.92 | 5.04 (0.84) | 4.81 |

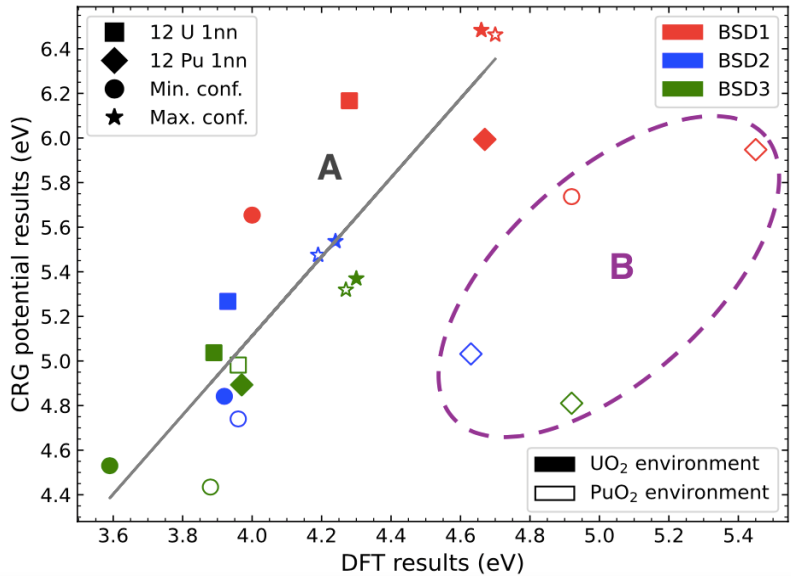

Figure 15: Comparison of the DFT+U results with the CRG potential calculation data of BSD formation energies. Solid (empty) markers represent the data obtained in the case of BSD-U (BSD-Pu) with a $UO_2$ ($PuO_2$) environment beyond the 1nn shell of the cation vacancy. Red, blue, and green markers are the calculated values of respectively BSD1, BSD2 and BSD3 formation energies. In gray, the linear regression performed on the data of the set A.

## VI. Conclusion

We presented a systematic study coupling interatomic potential and electronic structure calculations aimed at assessing quantitatively the influence of cationic chemical disorder on the bound Schottky defect (BSD) formation energy in $(U,Pu)O_2$. First, we proposed an original procedure to estimate the cutoff distance $R_D$ from which the chemical environment does not significantly affect the BSD formation energies based on CRG potential calculations. We found that $R_D \simeq 7.0$ Å, which corresponds to the first three cation shells around the BSD cation vacancy. Then, we performed a systematic exploration of the cation configurations up to the second nearest neighbor of the BSD cation vacancy, and provided for the first time a valuable database of tens of thousands of BSD formation energies in (U,Pu)$O_2$, computed with the CRG potential. This allowed for the evaluation of the maximum energy interval $\Delta E_F$ in which the BSD formation energy can range, depending on the composition of the surrounding local environment. In addition, a few targeted configurations were selected in order to benchmark the CRG results with the DFT+$U$ method. Our electronic structure calculations showed a rather good qualitative agreement with the CRG results for some configurations. We observed discrepancies in the case of a $Pu^{4+}$-rich local environment, but this might be due to a difficult convergence of some DFT+$U$ calculations. Nevertheless, in all cases a systematic mismatch is observed between CRG and DFT+$U$, most likely because of the scaling between the formation energies of $UO_2$ and $PuO_2$ yielded by the two methods. Thus, the coupled interatomic potential-DFT+U approach employed in the present study represents a valid strategy to study point defect properties in disordered materials such as (U,Pu)$O_2$.

This is the first extensive study on the effect of cationic chemical disorder on defect properties in (U,Pu)$O_2$ MOX compounds. The exploration strategy presented here can be applied to other kind of defects, such as oxygen or cation vacancies, in other types of disordered compounds (such as high-entropy alloys), as well as for the study of defect migration energies and prefactors to come in the calculation of self-diffusion coefficients. We show, however, that this approach is insufficient to explore the possible residual effect beyond the second shell. An exhaustive exploration in this case is clearly unfeasible due to the large number of configurations to be considered. This could be performed in future work using machine-learning methods. The database produced in the present study can serve as a training database for machine-learning energy models that could then be used to simulate the microstructure evolution of (U,Pu)$O_2$ MOX fuels under irradiation, in which the formation of point defects plays an important role.

## Declaration of competing interest

The authors declare that they have no known competing financial interests or personal relationships that could have appeared to influence the work reported in this paper.

## Data Availability Statement

The data that support the findings of this study are available from the corresponding author upon reasonable request.

## Acknowledgments

In the present work, high-performance resources from the Grand Equipement National de Calcul Intensif (GENCI) [Très Grand Centre de Calcul (TGCC) and Centre Informatique National de l'Enseignement Supérieur (CINES)] were used for the calculations. The authors wish to express their gratitude to S. Rajaonson, M. Karcz, G. Jomard and I. Cheik Njifon for the fruitful discussions. This research is part of the Investigations Supporting MOX Fuel licensing in ESNII Protocol Reactors (INSPYRE) Project, which has received funding from the Euratom research and training program 2014-2018 under grant 754329. This work was supported by the Cross-cutting basic research Program (RTA Program) of the CEA Energy Division.

# VII. References

[1] S. O. Vălu *et al.*, "The high-temperature heat capacity of the (Th,U)O$_2$ and (U,Pu)O$_2$ solid solutions," *J. Nucl. Mater.*, vol. 484, pp. 1–6, Feb. 2017, doi: 10.1016/j.jnucmat.2016.11.010.

[2] T. L. Markin and R. S. Street, "The uranium-plutonium-oxygen ternary phase diagram," *J. Inorg. Nucl. Chem.*, vol. 29, no. 9, pp. 2265–2280, Sep. 1967, doi: 10.1016/0022-1902(67)80281-1.

[3] N. H. Brett, "Oxidation products of plutonium dioxide-uranium dioxide solid solutions in air at 750 C," *J. Inorg. Nucl. Chem.*, vol. 28, no. 5, 1966.

[4] L. Lyon and W. E. Baily, "The solid-liquid phase diagram for the UO$_2$-PuO$_2$ system," *J. Nucl. Mater.*, vol. 22, p. 8, 1967.

[5] L. F. Epstein, "Ideal solution behavior and heats of fusion from the UO$_2$-PuO$_2$ phase diagram," *J. Nucl. Mater.*, vol. 22, no. 3, pp. 340–349, Jun. 1967.

[6] R. Böhler *et al.*, "Recent advances in the study of the UO$_2$–PuO$_2$ phase diagram at high temperatures," *J. Nucl. Mater.*, vol. 448, no. 1–3, pp. 330–339, May 2014, doi: 10.1016/j.jnucmat.2014.02.029.

[7] C. Guéneau *et al.*, "Thermodynamic modelling of advanced oxide and carbide nuclear fuels: Description of the U–Pu–O–C systems," *J. Nucl. Mater.*, vol. 419, no. 1–3, pp. 145–167, Dec. 2011, doi: 10.1016/j.jnucmat.2011.07.033.

[8] C. O. T. Galvin, P. A. Burr, M. W. D. Cooper, P. C. M. Fossati, and R. W. Grimes, "Using molecular dynamics to predict the solidus and liquidus of mixed oxides (Th,U)O$_2$, (Th,Pu)O$_2$ and (Pu,U)O$_2$," *J. Nucl. Mater.*, vol. 534, p. 152127, Jun. 2020, doi: 10.1016/j.jnucmat.2020.152127.

[9] P. Martin *et al.*, "XAS study of (U$_{1-y}$Pu$_y$)O$_2$ solid solutions," *J. Alloys Compd.*, vol. 444–445, pp. 410–414, Oct. 2007, doi: 10.1016/j.jallcom.2007.01.032.

[10] J. Wiktor, M.-F. Barthe, G. Jomard, M. Torrent, M. Freyss, and M. Bertolus, "Coupled experimental and DFT + U investigation of positron lifetimes in UO$_2$," *Phys. Rev. B*, vol. 90, no. 18, Nov. 2014, doi: 10.1103/PhysRevB.90.184101.

[11] R. Bès *et al.*, "Experimental evidence of Xe incorporation in Schottky defects in UO$_2$," *Appl. Phys. Lett.*, vol. 106, no. 11, p. 114102, Mar. 2015, doi: 10.1063/1.4914300.

[12] E. Vathonne, "Étude par calcul de structure électronique des dégâts d'irradiation dans le combustible nucléaire UO$_2$: comportement des défauts ponctuels et gaz de fission," p. 271.

[13] A. E. Thompson and C. Wolverton, "First-principles study of noble gas impurities and defects in UO$_2$," *Phys. Rev. B*, vol. 84, no. 13, Oct. 2011, doi: 10.1103/PhysRevB.84.134111.

[14] H. Matzke, "Atomic Transport Properties in UO$_2$ and Mixed Oxides (U,Pu)O$_2$," *J Chem Soc Faraday Trans 2*, vol. 83, no. 2, pp. 1121–1142, 1983.

[15] I. Cheik Njifon, "Modélisation des modifications structurales, électroniques et thermodynamiques induites par les défauts ponctuels dans les oxydes mixtes à base d'actinides (U,Pu)O2," Aix-Marseille Université, 2018.

[16] V. I. Anisimov, J. Zaanen, and O. K. Andersen, "Band theory and Mott insulators: Hubbard *U* instead of Stoner *I*," *Phys. Rev. B*, vol. 44, no. 3, pp. 943–954, Jul. 1991, doi: 10.1103/PhysRevB.44.943.

[17] J. von Pezold, A. Dick, M. Friák, and J. Neugebauer, "Generation and performance of special quasirandom structures for studying the elastic properties of random alloys: Application to Al-Ti," *Phys. Rev. B*, vol. 81, no. 9, Mar. 2010, doi: 10.1103/PhysRevB.81.094203.

[18] A. Zunger, S.-H. Wei, L. G. Ferreira, and J. E. Bernard, "Special quasirandom structures," *Phys. Rev. Lett.*, vol. 65, no. 3, pp. 353–356, Jul. 1990, doi: 10.1103/PhysRevLett.65.353.

[19] H. Balboa, L. Van Brutzel, A. Chartier, and Y. Le Bouar, "Damage characterization of (U,Pu)O2 under irradiation by molecular dynamics simulations," *J. Nucl. Mater.*, vol. 512, pp. 440–449, 2018, doi: 10.1016/j.jnucmat.2018.07.056.

[20] M. W. D. Cooper, M. J. D. Rushton, and R. W. Grimes, "A many-body potential approach to modelling the thermomechanical properties of actinide oxides," *J. Phys. Condens. Matter*, vol. 26, no. 10, p. 105401, Mar. 2014, doi: 10.1088/0953-8984/26/10/105401.

[21] C. Takoukam-Takoundjou, E. Bourasseau, and V. Lachet, "Study of thermodynamic properties of $U_{1-y}Pu_yO_2$ MOX fuel using classical molecular Monte Carlo simulations," *J. Nucl. Mater.*, vol. 534, p. 152125, Jun. 2020, doi: 10.1016/j.jnucmat.2020.152125.

[22] X.-Y. Liu and D. A. Andersson, "Small uranium and oxygen interstitial clusters in $UO_2$: An empirical potential study," *J. Nucl. Mater.*, vol. 547, p. 152783, Apr. 2021, doi: 10.1016/j.jnucmat.2021.152783.

[23] K. Li, C. C. Fu, M. Nastar, F. Soissons, and M. Y. Lavrentiev, "Magnetochemical effects on phase stability and vacancy formation in fcc Fe-Ni alloys," *Arxiv220304688 Cond-Matmtrl-Sci*.

[24] L.-F. Wang, B. Sun, H.-F. Liu, D.-Y. Lin, and H.-F. Song, "Thermodynamics and kinetics of intrinsic point defects in plutonium dioxides," *J. Nucl. Mater.*, vol. 526, p. 151762, Dec. 2019, doi: 10.1016/j.jnucmat.2019.151762.

[25] M. Nastar and F. Soisson, "Atomistic modeling of phase transformations: Point-defect concentrations and the time-scale problem," *Phys. Rev. B*, vol. 86, no. 22, p. 220102, Dec. 2012, doi: 10.1103/PhysRevB.86.220102.

[26] G. Kresse and J. Furthmüller, "Efficient iterative schemes for *ab initio* total-energy calculations using a plane-wave basis set," *Phys. Rev. B*, vol. 54, no. 16, pp. 11169–11186, Oct. 1996, doi: 10.1103/PhysRevB.54.11169.

[27] G. Kresse and J. Hafner, "*Ab initio* molecular-dynamics simulation of the liquid-metal–amorphous-semiconductor transition in germanium," *Phys. Rev. B*, vol. 49, no. 20, pp. 14251–14269, May 1994, doi: 10.1103/PhysRevB.49.14251.

[28] P. E. Blöchl, "Projector augmented-wave method," *Phys. Rev. B*, vol. 50, no. 24, pp. 17953–17979, Dec. 1994, doi: 10.1103/PhysRevB.50.17953.

[29] G. Kresse and D. Joubert, "From ultrasoft pseudopotentials to the projector augmented-wave method," *Phys. Rev. B*, vol. 59, no. 3, pp. 1758–1775, Jan. 1999, doi: 10.1103/PhysRevB.59.1758.

[30] J. Klimeš, D. R. Bowler, and A. Michaelides, "Chemical accuracy for the van der Waals density functional," *J. Phys. Condens. Matter*, vol. 22, no. 2, p. 022201, Jan. 2010, doi: 10.1088/0953-8984/22/2/022201.

[31] A. I. Liechtenstein, V. I. Anisimov, and J. Zaanen, "Density-functional theory and strong interactions: Orbital ordering in Mott-Hubbard insulators," *Phys. Rev. B*, vol. 52, no. 8, pp. R5467–R5470, Aug. 1995, doi: 10.1103/PhysRevB.52.R5467.

[32] I. C. Njifon, M. Bertolus, R. Hayn, and M. Freyss, "Electronic Structure Investigation of the Bulk Properties of Uranium–Plutonium Mixed Oxides (U, Pu)$O_2$," *Inorg. Chem.*, vol. 57, no. 17, pp. 10974–10983, Sep. 2018, doi: 10.1021/acs.inorgchem.8b01561.

[33] M. S. Talla Noutack, G. Jomard, M. Freyss, and G. Geneste, "Structural, electronic and energetic properties of uranium–americium mixed oxides $U_{1-y}Am_yO_2$ using DFT+*U* calculations," *J. Phys. Condens. Matter*, vol. 31, no. 48, p. 485501, Dec. 2019, doi: 10.1088/1361-648X/ab395e.

[34] G. Jomard, B. Amadon, F. Bottin, and M. Torrent, "Structural, thermodynamic, and electronic properties of plutonium oxides from first principles," *Phys. Rev. B*, vol. 78, no. 7, p. 075125, Aug. 2008, doi: 10.1103/PhysRevB.78.075125.

[35] S. B. Wilkins *et al.*, "Direct observation of electric-quadrupolar order in UO$_2$," *Phys. Rev. B*, vol. 73, no. 6, Feb. 2006, doi: 10.1103/PhysRevB.73.060406.

[36] P. Santini, S. Carretta, G. Amoretti, R. Caciuffo, N. Magnani, and G. H. Lander, "Multipolar interactions in f -electron systems: The paradigm of actinide dioxides," *Rev. Mod. Phys.*, vol. 81, no. 2, pp. 807–863, Jun. 2009, doi: 10.1103/RevModPhys.81.807.

[37] B. Dorado and P. Garcia, "First-principles DFT + U modeling of actinide-based alloys: Application to paramagnetic phases of UO 2 and (U,Pu) mixed oxides," *Phys. Rev. B*, vol. 87, no. 19, p. 195139, May 2013, doi: 10.1103/PhysRevB.87.195139.

[38] G. Raphael and R. Lallement, "Susceptibilité magnétique de PuO$_2$," *Solid State Commun.*, vol. 6, no. 6, 1968.

[39] S. Kern, C.-K. Loong, G. L. Goodman, B. Cort, and G. H. Lander, "Crystal-field spectroscopy of PuO $_2$ : further complications in actinide dioxides," *J. Phys. Condens. Matter*, vol. 2, no. 7, pp. 1933–1940, Feb. 1990, doi: 10.1088/0953-8984/2/7/024.

[40] Y. Tokunaga *et al.*, "NMR studies of actinide dioxides," *J. Alloys Compd.*, vol. 444–445, pp. 241–245, Oct. 2007, doi: 10.1016/j.jallcom.2007.03.082.

[41] S. Kern *et al.*, "Crystal-field transition in PuO 2," *Phys. Rev. B*, vol. 59, no. 1, pp. 104–106, Jan. 1999, doi: 10.1103/PhysRevB.59.104.

[42] J. T. Pegg, A. E. Shields, M. T. Storr, A. S. Wills, D. O. Scanlon, and N. H. de Leeuw, "Hidden magnetic order in plutonium dioxide nuclear fuel," *Phys. Chem. Chem. Phys.*, vol. 20, no. 32, pp. 20943–20951, 2018, doi: 10.1039/C8CP03583K.

[43] B. Dorado, G. Jomard, M. Freyss, and M. Bertolus, "Stability of oxygen point defects in UO$_2$ by first-principles DFT + U calculations: Occupation matrix control and Jahn-Teller distortion," *Phys. Rev. B*, vol. 82, no. 3, Jul. 2010, doi: 10.1103/PhysRevB.82.035114.

[44] R. P. Feynman, "Forces in Molecules," *Phys. Rev.*, vol. 56, no. 4, pp. 340–343, Aug. 1939, doi: 10.1103/PhysRev.56.340.

[45] G. H. F. Diercksen and S. Wilson, Eds., *Methods in Computational Molecular Physics*. Dordrecht: Springer Netherlands, 1983. doi: 10.1007/978-94-009-7200-1.

# Appendix A: Limitations of the traditional formation energy formula for disordered solutions

We demonstrate in this section that the traditional formula (Eq. 2.1) is not well suited for the formation energy of vacancy-type defects in disordered solutions, applied in particular to the case of a BSD defect in (U,Pu)O$_2$. By definition, the BSD formation energy is given by:

$$\hat{E}_f^X = E_D - E_R^X + \hat{E}_{\text{norm}} \quad (A1)$$

where $X = $ U, Pu is the chemical species that has been removed to form the cation vacancy. The normalization term, which is necessary to account for the varying number of atoms between the reference supercell (with energy $E_R^X$) and the defected one (with energy $E_D$), is:

$$\hat{E}_{\text{norm}} = \mu_X + 2\mu_O \quad (A2)$$

On the other hand, rearranging of Eq. 2.1 yields:

$$E_f^X = E_D - E_R^X + E_{\text{norm}} \quad (A3)$$

where the normalization term is given by:

$$E_{\text{norm}} = \frac{E_R^X}{N} \quad (A4)$$

Since the energy of a reference supercell characterized by a Pu concentration $y$ is given by definition by:

$$E_R^X = N[(1-y)\mu_U + y\mu_{\text{Pu}} + 2\mu_O] \quad (A5)$$

the normalization term of Eq. 2.1 can be also written as:

$$E_{\text{norm}} = (1-y)\mu_U + y\mu_{\text{Pu}} + 2\mu_O = \mu_U + 2\mu_O + y(\mu_{\text{Pu}} - \mu_U) \quad (A6)$$

Therefore, as opposed to the correct reference $\mu_X + 2\mu_O$ (Eq. A2), in Eq. 2.1 the formation energy is obtained with respect to an energy reference where the chemical potentials of U and Pu are mixed proportionally to the concentration of each species. Besides being physically questionable, the consequence is that the energy reference depends on the Pu concentration $y$, and varying $y$ introduces an artificial shift of formation energy that is intertwined with the effect of cationic disorder.

In conclusion, Eq. 2.1 provides a wrong energy reference for $E_f^X$, and does not allow to compare between supercells with different values of $y$.

# Appendix B: General BSD formation energy formula, and simplification for ideal solutions

In a disordered binary compound, the fact that several host species can be chosen to replace the vacancy leads to two different definitions of formation energy for the same defect. Indeed, in the case of (U,Pu)O₂, the BSD formation energy can be defined in two different ways, depending on whether the reference supercell contains a U or a Pu atom in place of the cation vacancy, as is the case in Eq. 2.1 or Eq. 2.2. In the general case of non-ideal solution, the two definitions are:

$$E_f^U = E_D - E_R^U + \mu_U + 2\mu_O - k_B T \log(1-y) \quad (B1)$$

$$E_f^{\text{Pu}} = E_D - E_R^{\text{Pu}} + \mu_{\text{Pu}} + 2\mu_O - k_B T \log(y) \quad (B2)$$

The difference between the two equations above lies simply in a different choice of the energy reference, which in fact does not affect the equilibrium defect concentration given by the following ensemble average:

$$C_{\text{BSD}} = \left\langle \exp\left(-\frac{E_f^U}{k_B T}\right) \right\rangle = \left\langle \exp\left(-\frac{E_f^{\text{Pu}}}{k_B T}\right) \right\rangle \quad (B3)$$

Therefore, choosing either one of the two equations is equally valid to analyze the impact of chemical disorder on the defect formation energy, since the only difference between $E_f^U$ and $E_f^{Pu}$ is a shift in the energy reference.

In the case of an ideal solution, where any local-order effect can be neglected, it can be shown [23] that the chemical potentials $\mu_U^0$ and $\mu_{Pu}^0$ are given by:

$$\mu_U^0 + 2\mu_O^0 = \frac{E(UO_2)}{N} + k_B T \log(1-y) \quad (B4)$$

$$\mu_{Pu}^0 + 2\mu_O^0 = \frac{E(PuO_2)}{N} + k_B T \log(y) \quad (B5)$$

where $E(UO_2)$ and $E(PuO_2)$ are the energies of the pure undefected UO$_2$ and PuO$_2$ supercells. The BSD formation energies in the ideal solution $E_f^{U,0}$ and $E_f^{Pu,0}$ are then:

$$E_f^{U,0} = E_D - E_R^U + \frac{E(UO_2)}{N} \quad (B6)$$

$$E_f^{Pu,0} = E_D - E_R^{Pu} + \frac{E(PuO_2)}{N} \quad (B7)$$

In the present work, we rely on Eq. B6, and use $E_f^{U,0}$ to analyze the impact of chemical disorder on the defect formation energy, and do not consider $E_f^{Pu,0}$ as we assume that it would yield the same defect concentration $C_{BSD}$. However, this assumption holds only as long as short-range order is negligible. In the opposite case, the equality:

$$C_{BSD} = \left\langle \exp\left(-\frac{E_f^{U,0}}{k_B T}\right) \right\rangle = \left\langle \exp\left(-\frac{E_f^{Pu,0}}{k_B T}\right) \right\rangle \quad (B8)$$

would be no longer true. Future works will be dedicated to assess any possible effects of short-range order on the BSD properties in (U,Pu)O$_2$.

## Appendix C: Estimation of the cutoff distance

Table 6 shows the main characteristics of the composite supercells generated for the study presented in Section III. We remind that a composite supercell $C_n$ is generated by modifying locally the cationic chemical environment of the cation vacancy of bound Schottky defects up to the $n^{th}$ shell. The knowledge of the radius of each shell is essential for our purpose because we rely on this quantity to determine the cutoff distance beyond which the cationic chemical disorder has no significant effect on defect formation energies in (U,Pu)O$_2$.

Figure 16 shows the Pu nominal concentration in the generated composite supercells, which ranges between 49% and 51%. Hence, we observe a slight deviation of the nominal chemical composition of the composite supercell with respect to the one in the reference supercells presented in Section III, i.e., SQS and ordered supercells, that contains 50% Pu. According to

Figure 16, the variation of the Pu concentration tends to increase from supercells $C_1$ to $C_8$, which is explained by the increasing number of cations replaced, which raises the probability to modify the Pu nominal concentration. In SQS supercells (i.e., disordered supercells) with a nominal Pu concentration of 50 %, the Pu concentration in each shell is not exactly equal to 50%. As a matter of fact, we notice large deviations of the Pu nominal concentration in the cases of supercells $C_7$ in Figure 16, because 48 cations are replaced to generate the composite supercells $C_7$ according to Table 6, which is a far larger number than in the case of the other composite supercells.

Furthermore, we show in Figure 17 and Figure 18 the results obtained in the case of the BSD1 and BSD2 respectively, concerning the impact of the local environment composition presented in Section III.2. The conclusions are the same than the ones obtained in the case of BSD3 (see Figure 5), and discussed in Section III.2.

Table 6: Features of the eighth first cation coordination shells around a cation in (U,Pu)O$_2$, associated to the composite supercells from $C_1$ to $C_8$.

| Coordination shell | Number of cations on the shell | Radius of the shell as a function of the lattice parameter $a$ | Radius (in Å) of the shell with $a = 5.43$ Å | Composite supercell |
|---|---|---|---|---|
| 1$^{rst}$ | 12 | $\frac{\sqrt{2}}{2}a$ | 3.84 | $C_1$ |
| 2$^{nd}$ | 6 | $a$ | 5.43 | $C_2$ |
| 3$^{rd}$ | 24 | $\sqrt{\frac{3}{2}}a$ | 6.65 | $C_3$ |
| 4$^{th}$ | 12 | $\sqrt{2}a$ | 7.68 | $C_4$ |
| 5$^{th}$ | 24 | $\frac{\sqrt{10}}{2}a$ | 8.59 | $C_5$ |
| 6$^{th}$ | 8 | $\sqrt{3}a$ | 9.41 | $C_6$ |
| 7$^{th}$ | 48 | $\frac{\sqrt{14}}{2}a$ | 10.16 | $C_7$ |
| 8$^{th}$ | 6 | $2a$ | 10.86 | $C_8$ |
| $n^{th}$ | - | $\frac{\sqrt{2n}}{2}a$ | - | $C_n$ |

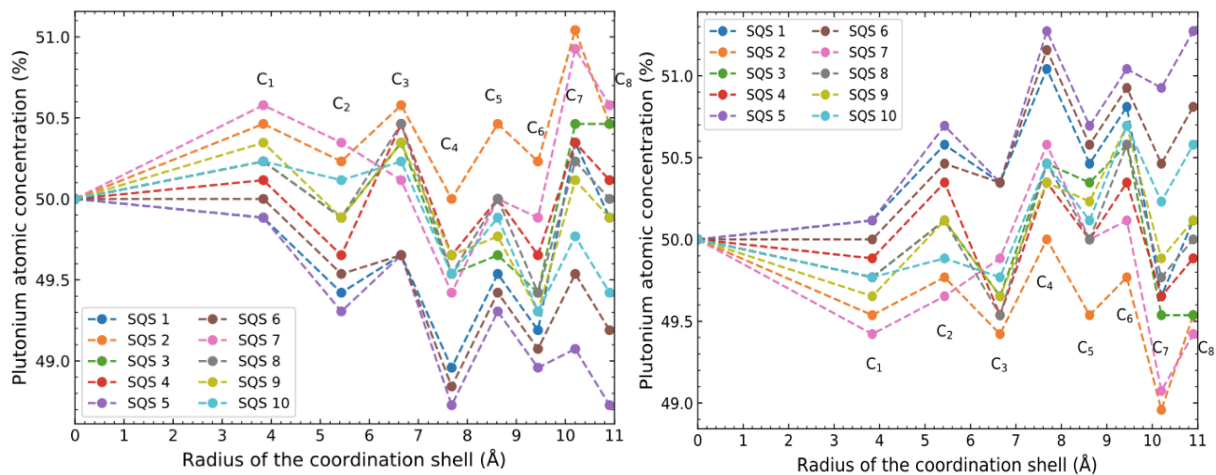

*Figure 16: Pu atomic concentration in composite supercells generated by a) "implanting ordered clusters in SQS supercells" and by b) "implanting SQS clusters in ordered supercells". Each color refers to the set of composite supercells generated with the ordered supercell plane-1 and a given SQS supercell.*

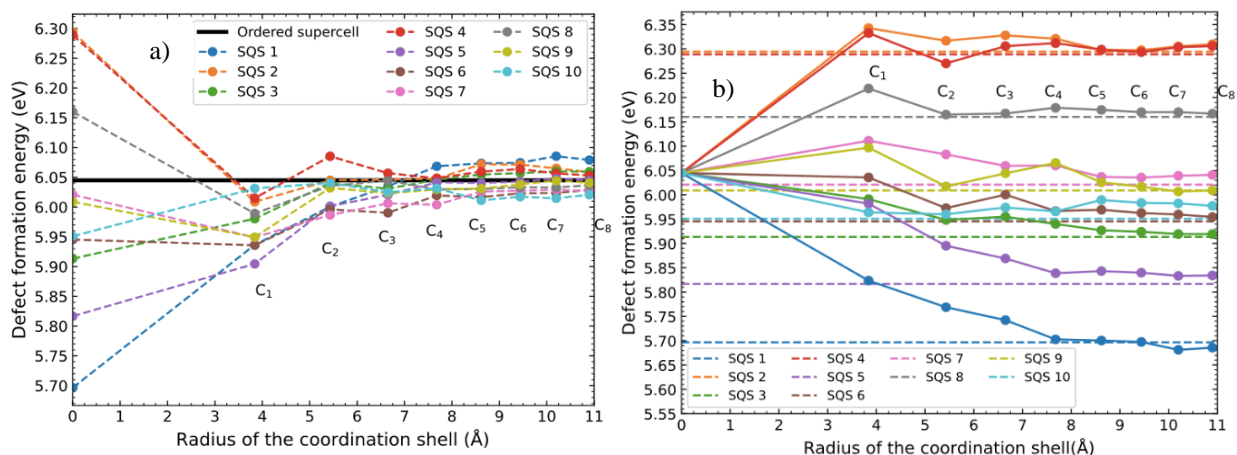

*Figure 17: BSD1 formation energy calculated in composite supercells generated by a) "implanting ordered clusters in SQS supercells" and by b) "implanting SQS clusters in ordered supercells". In a) the black line refers to the BSD3 formation energy calculated in the ordered supercell plane 1. In b) the dashed lines refer to the BSD3 formation energy in each of the ten SQS supercells.*

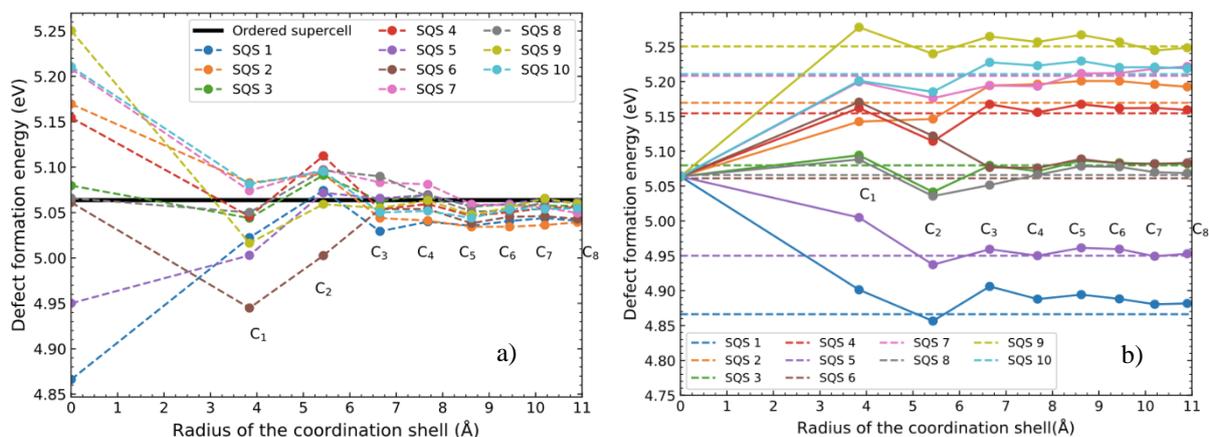

*Figure 18: BSD2 formation energy calculated in composite supercells generated by a) "implanting ordered clusters in SQS supercells" and by b) "implanting SQS clusters in ordered supercells". In a) the black line refers to the BSD3 formation energy*

*calculated in the ordered supercell plane 1. In b) the dashed lines refer to the BSD3 formation energy in each of the ten SQS supercells.*

# Appendix D: Systematic evaluation

Figure 19a (resp. Figure 19b, Figure 19c) compares the results of the systematic study by means of interatomic-potential calculations for the BSD1-U, BSD2-U, and BSD3-U obtained with a Pu nominal content of 50 % (resp. 75%, 100%) in the outer region, i.e., beyond the first coordination shell of the cation vacancy. The results are similar to the ones presented in the case of 0 % and 25 % in Section IV.2.1. Figure 20a, Figure 20b, and Figure 20c shows the results obtained in the cases of BSD-Pu. As we admit in Section II and in Appendix C, there are no significant differences qualitatively in the results obtained in the cases of BSD-U and BSD-Pu, which explain that we could reasonably consider the unique case of the BSD-U in Section IV to study the effect of cationic chemical disorder on defect formation energy in $(U,Pu)O_2$. We observe indeed a slight shift of the BSD formation energies between the BSD-U (Figure 19) and BSD-Pu (Figure 20) cases.

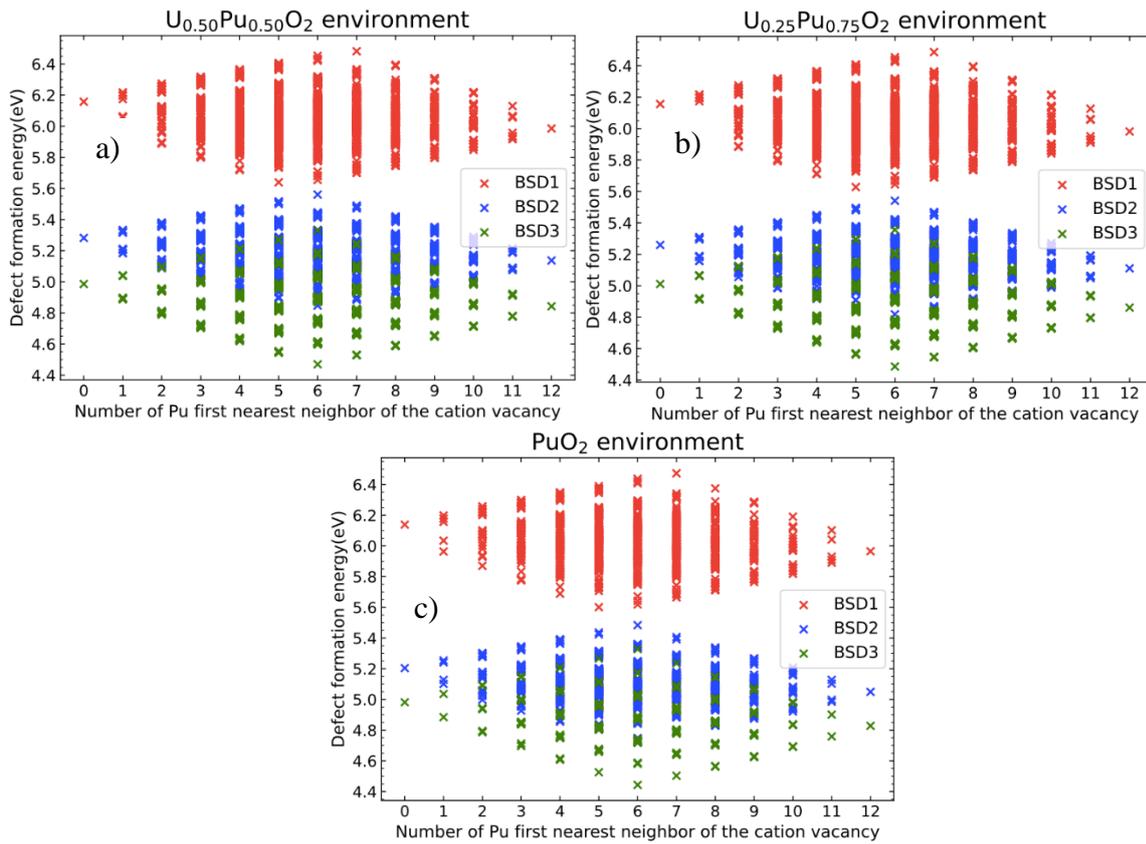

*Figure 19: BSD1-U (red), BSD2-U (blue) and BSD3-U (red) formation energies calculated using the CRG potential with (a) 25% Pu, (b) 75% Pu and (c) 100 % Pu on the cation sublattice beyond the first coordination shell.*

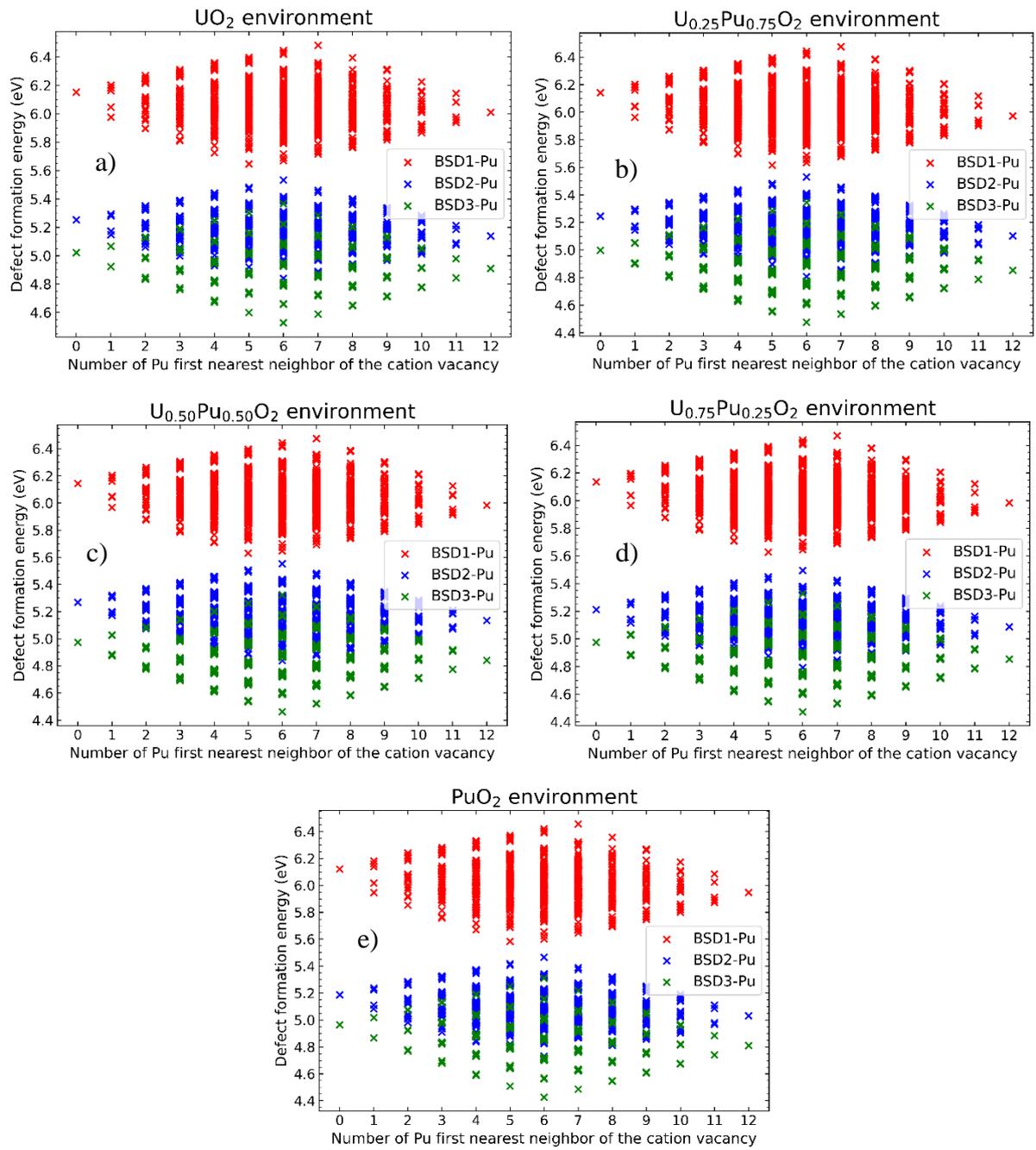

*Figure 20: BSD1-Pu (red), BSD2-Pu (blue) and BSD3-Pu (red) formation energies calculated using the CRG potential with (a) 0% Pu, (b) 25% Pu, (c) 50 % Pu, (d) 75 % Pu, and (d) 100 % Pu on the cation sublattice beyond the first coordination shell.*